\documentclass[hidelinks,12pt]{article}
\usepackage{epsfig,amsfonts,amssymb}
\pdfoutput=1
\usepackage{comment}
\input epsf.sty
\usepackage{cite}
\usepackage{epsfig,amssymb,euscript,xspace,xcolor}
\usepackage{amsmath,mathtools,empheq,amsthm,hyperref,graphicx}
\usepackage{mathrsfs,float}  
\usepackage{pgf,tikz,pgfplots}
\usetikzlibrary{arrows}
\usepackage[T1]{fontenc} 
\usepackage{tikz,caption,subcaption,marvosym} 
\usetikzlibrary{decorations.markings,arrows,snakes}
\usepackage[skip=10pt]{caption} 
\usepackage{comment}

\usepackage[utf8]{inputenc}

\usepackage[titles]{tocloft}

\definecolor{lightblue}{rgb}{.1,.4,.5}
\definecolor{brown1}{rgb}{.64,.43,.138}

\usepackage{hyperref,cite}
\hypersetup{linktocpage, colorlinks=true,linkcolor= blue,citecolor=blue,urlcolor=lightblue}

\def\ZZZ{{\hbox{ Z\kern-1.6mm Z}}}
\def\RRR{{\hbox{ R\kern-2.4mm R}}}
\def\CCC{{\hbox{ C\kern-2.0mm C}}}
\def\zzz{{\hbox{z\kern-1mm z}}}

\newcommand{\qeq}{{\hbox{=\kern-2.3mm ? \kern.5mm }}}
\renewcommand{\qeq}{=}

\newcommand{\be}{\begin{equation}}
\newcommand{\ee}{\end{equation}}
\newcommand{\ben}{\begin{eqnarray}\displaystyle}
\newcommand{\een}{\end{eqnarray}}

\def\one{{\hbox{ 1\kern-.8mm l}}}
\def\zero{{\hbox{ 0\kern-1.5mm 0}}}

\renewcommand{\theequation}{\thesection.\arabic{equation}}
\renewcommand{\theequation}{\arabic{equation}}

\newcommand{\bea}[1]{\begin{eqnarray}\label{#1} }
\newcommand{\eea}{\end{eqnarray}}

\usepackage{bm}

\def\bea{\begin{eqnarray}}
\def\eea{\end{eqnarray}}
\def\be{\begin{equation}}
\def\ee{\end{equation}}

\definecolor{wvvxds}{rgb}{0.396078431372549,0.3411764705882353,0.8235294117647058}
\definecolor{dbwrru}{rgb}{0.8588235294117647,0.3803921568627451,0.0784313725490196}
\definecolor{dtsfsf}{rgb}{0.8274509803921568,0.1843137254901961,0.1843137254901961}
\definecolor{wrwrwr}{rgb}{0.3803921568627451,0.3803921568627451,0.3803921568627451}
\definecolor{cqcqcq}{rgb}{0.7529411764705882,0.7529411764705882,0.7529411764705882}
\definecolor{rvwvcq}{rgb}{0.08235294117647059,0.396078431372549,0.7529411764705882}


\begin{document}

\baselineskip 24pt

\begin{center}

{\Large \bf  Stokes Polytopes : The positive geometry for $\phi^{4}$ interactions}

\end{center}

\vskip .5cm
\medskip

\vspace*{4.0ex}

\baselineskip=18pt

\centerline{\large \rm Pinaki Banerjee$^{s}$, Alok Laddha$^{t}$ and Prashanth Raman$^{u}$}

\vspace*{4.0ex}

\centerline{ \it ~$^s$International Centre for Theoretical Sciences,}
\centerline{ \it Tata Institute of Fundamental Research}
\centerline{ \it Shivakote, Bengaluru 560 089, India} 

\vspace*{1.0ex}

\centerline{\it ~$^t$Chennai Mathematical Institute,  SIPCOT IT Park, Siruseri, Chennai, 603103 India} 

\vspace*{1.0ex}

\centerline{\it ~$^u$Institute of Mathematical Sciences, Taramani, Chennai 600 113, India}
\centerline{\it ~$^u$Homi Bhabha National Institute, Anushakti Nagar, Mumbai 400085, India}

\vspace*{1.0ex}
\centerline{\small \textit{E-mail :}  \href{mailto:pinaki.banerjee@icts.res.in}{ \texttt{pinaki.banerjee@icts.res.in}}, \href{mailto:aladdha@cmi.ac.in}{\texttt{aladdha@cmi.ac.in}},  \href{mailto:prashanthr@imsc.res.in}{\texttt{prashanthr@imsc.res.in}}}

\vspace*{5.0ex}

\centerline{\bf Abstract} \bigskip

In a remarkable recent work \cite{Arkani-Hamed:2017mur}, the amplituhedron program was extended to  the realm of non-supersymmetric scattering amplitudes. In particular it was shown that for  tree-level planar diagrams in massless $\phi^{3}$ theory (and its close cousin, bi-adjoint $\phi^{3}$ theory) a polytope known as the associahedron sits inside the kinematic space and is the amplituhedron for the theory. Precisely as in the case of amplituhedron,  it was shown that scattering amplitude can be obtained from the canonical form  associated to the Associahedron. Combinatorial and geometric properties of associahedron naturally encode properties like locality and unitarity of (tree level) scattering amplitudes.  In this paper we attempt to extend this program to planar amplitudes in massless $\phi^{4}$ theory. 
We show that tree-level planar amplitudes in this theory can be obtained from geometry of objects known as the Stokes polytope which sits naturally inside the kinematic space. As in the case of associahedron we show that the canonical form on these Stokes polytopes can be used to compute scattering amplitudes for quartic interactions. However unlike associahedron, Stokes polytope of a given dimension is not unique and as we show, one must sum over all of them to obtain the complete scattering amplitude. Not all Stokes polytopes contribute equally and we argue that the corresponding weights depend on purely combinatorial  properties of the Stokes polytopes.  As in the case of $\phi^{3}$ theory, we show how factorization of Stokes polytope implies unitarity and locality of the amplitudes.

\vfill \eject

\baselineskip=18pt

\tableofcontents

\section{Introduction}

In \cite{Arkani-Hamed:2017mur}, authors extended the ``amplituhedron program'' \cite{Arkani-Hamed:2013jha} of analysing scattering amplitudes in super-symmetric quantum field theories to a class of non-supersymmetric theories.  In particular, for tree level planar diagrams  in massless $\phi^{3}$ theory (or it's close cousin, all tree level diagrams in bi-adjoint scalar field theory) a precise connection was established between so-called planar scattering form on kinematic space, a polytope known as associahedron and tree-level scattering amplitudes. Fascinating attempts have also been made to  extend the program to 1-loop amplitudes in $\phi^{3}$ theory, where the corresponding polytope is an object already known to mathematicians known as Halohedron \cite{Salvatori:2018fjp, Salvatori:2018aha}. 

This work has far reaching ramifications for our understanding of scattering amplitudes. Specifically, two new perspectives has emerged : 
\begin{enumerate}
\item Understanding of amplitudes not as functions but as differential forms on kinematic space,  
\item A precise connection between these forms and polytopes located inside the kinematic space. This new perspective leads one to a new understanding of locality , unitarity and various other properties (like soft limits and recursion relations) of scattering amplitudes from combinatorial and geometric properties of the polytopes. 
\end{enumerate}

Another beautiful result was established in \cite{Arkani-Hamed:2017mur} that gave a new understanding of the formulae of Cachazo, He and Yuan (CHY) for tree-level scattering amplitudes \cite{Cachazo:2013hca}. The CHY formula expresses scattering amplitude for a large class of theories (including planar diagrams in massless $\phi^{3}$ theory) as integrals over certain world-sheet moduli space \cite{Cachazo:2013hca}. It has been known for some time that compactification of this moduli space is  an associahedron \cite{Devadoss, Deligne}.  In \cite{Arkani-Hamed:2017mur} it was shown that this ``worldsheet associahedron'' is in fact diffeomorphic to the associahedron sitting inside kinematic space! Scattering equations which are basic building blocks of CHY formula are precisely these diffeomorphisms. Whence it naturally followed that the CHY integrand for $\phi^{3}$ theory is a pullback of the  canonical scattering form on the associahedron.

This relationship between polytopes in kinematic space with CHY integrand however presents a puzzle. CHY formulae exist for (tree-level) amplitudes in a wide class of quantum field theories including planar diagrams in scalar field theories with $\phi^{p},\ p\ > 3$ interactions \cite{Baadsgaard:2015ifa, Baadsgaard:2016fel}. Thus it is a natural question to ask if for such theories, the CHY formula can also be understood in terms of differential forms and polytopes in kinematic space, with scattering equations defining the diffeomorphism. But before answering this question, we need to understand how to extend the ``amplituhedron program'' to such theories. In this paper, we take a small step in answering this second question in the context of quartic interactions.  

That is, we would like to ask if there is a relationship between (tree-level, planar) amplitudes in massless $\phi^{4}$ theory, scattering forms and polytopes in kinematic space.  As we show below, the answer is in the affirmative, although it differs from the idea of a single polytope such as associahedron which contains complete information about scattering amplitudes in several respects. 

We begin our analysis by trying to generalise one of the key observations of \cite{Arkani-Hamed:2017mur}, namely existence of a unique differential form on the kinematic space. Uniqueness of this form is however tied to a striking property of $\phi^{3}$ amplitudes called projectivity. Essentially projectivity captures the idea that planar amplitudes in massless $\phi^{3}$ theory have no pole at infinity in the kinematic space. However, from the days of BCFW \cite{Britto:2005fq} recusion relations \cite{Feng:2009ei}, it is well known that tree-level amplitudes for $\phi^{4}$ theory do have a pole at infinity and hence projectivity \emph{cannot} be used to define a unique differential form in this case. Although this looks like a formidable obstacle, there is a rather natural solution to the problem. As we show in section \ref{locat=n6section}, in the case of $n$-particle scattering, there is a family of \emph{unique} scattering forms in kinematic space, parametrised by quadrangulations $Q$ of a polygon\footnote{By quadrangulation we mean, splitting a polygon into quadrilaterals.} with $n$-vertices. Although no single form contains information about all the poles of the $n$ particle amplitude, the entire family of scattering forms do. For each of these forms parametrised by $Q$, a picture closely analogous to the picture in \cite{Arkani-Hamed:2017mur} emerges. 

As we show in section \ref{locat=n6section}, for each $Q$ of a hexagon, a  one dimensional positive geometry sits inside kinematic space of $n$ particles. It turns out that this positive geometry is a convex realisation of a specific  Stokes polytope. Stokes polytopes are combinatorial polytopes discovered by Baryshnikov in \cite{Baryshnikov}. Compared to the  associahedron which was discovered by Jim Stasheff in 60's \cite{Stasheff1,Stasheff2}, these polytopes were discovered rather recently in the context of studying singularities of quadratic forms. Convex realisations of the Stokes polytopes have been studied in \cite{Baryshnikov, chapoton, Palu}. As a convex realisation of the Stokes polytope will be relevant for us in the study of scattering amplitudes, we denote both the Stokes polytopes as well as their realisations as positive geometries as ${\cal S}^{Q}_{n}$.

For each of these Stokes polytopes ${\cal S}^{Q}_{n}$ whose dimension depends on $n$ and are paramterised by $Q$, the scattering form\footnote{It is worth mentioning that we need to distinguish between combinatorial polytopes like Associahedron and their convex realisations. A combinatorial polytope should be thought of as an abstract set of faces and incidence relations described in terms of some combinatorical data (\emph{e.g}, triangulations or quadrangulations). On the other hand, a convex realisation is the intersection of half-spaces defined by the positivity of some linear functions. A convex polytope is an example of a positive geometry \cite{Arkani-Hamed:2017tmz}. To a positive geometry, it is possible to associate a unique differential form, known as canonical form. In this article by polytopes we always mean convex polytopes.} descends to a unique canonical form with logarithmic singularities on the boundaries. As in the case of associahedron and $\phi^{3}$ amplitudes, this canonical form can be used to obtain $n$-particle planar scattering amplitude of the theory. However there is a key difference with the associahedron picture.  The form associated to a single polytope only yields some of the channel-contributions in such a way that a weighted sum over the polytopes produces complete amplitude ${\cal M}_{n}$. 

Our proposal for scattering amplitude obtained from combinatorial geometry of Stokes polytopes can be summarised by the formula

\begin{equation}
{\cal M}_{n}\ =\ \sum_{Q}\alpha_{Q} \, m_{n}(Q)
\end{equation}

where $m_{n}(Q)$ is the rational canonical function \cite{Arkani-Hamed:2017mur} associated to the form $\omega_{n}^{Q}$  and the weights  $\alpha_{Q}$ only depend on certain combinatorial properties of the quadrangulation $Q$ (see section \ref{primitives}). Although we do not have a  analytical formula for $\alpha_{Q}$ for arbitrary $n$, we check the validity of our proposal in a few examples. 

In section \ref{factorization}, we show that exactly as in the case of associahedron and $\phi^{3}$ theory, factorization properties of Stokes polytope imply the on-shell factorization of scattering amplitudes. A massless $\phi^{4}$ theory can be obtained from a theory of two scalar fields with cubic interaction where one of the (massive) fields is integrated out. In section \ref{cubicquartic}, we try to understand this connection in terms of polytopes and differential forms and argue that the combinatorial geometry of single Stokes polytope can not be derived from the geometry associated to cubic couplings. We end with conclusions.

\section{Planar scattering form and associahedron}

In this section, we summarise the key results of \cite{Arkani-Hamed:2017mur}. We review the construction of planar scattering form and kinematic associahedron for planar (tree-level) amplitudes in massless $\phi^{3}$ theory. For more details, we refer the reader to \cite{Arkani-Hamed:2017mur}. Throughout the paper, by amplitude we always mean reduced amplitude where momentum conserving $\delta$-function have been projected out.

\subsection{Kinematic space}
Kinematic space ($\mathcal{K}_n$) of $n$-massless momenta $p_i$ where $i =1,2, \ldots n$ is spanned by  $\binom n2$ number of Mandelstam variables,
\begin{align}
s_{ij} = (p_i + p_j)^2 = 2 p_i . p_j
\end{align} 
For spacetime dimensions $d < n-1$, all of them are not linearly independent and they need to satisfy the following condition
\begin{align}
\sum_{j=1; j \ne i}^{n}s_{ij} = 0, \quad i=1,2, \ldots n
\end{align} 
Thus the dimensionality of the kinematic space ($\mathcal{K}_n$) of $n$ massless particles reduces to
\begin{align}
dim (\mathcal{K}_n) = \binom n2 - n = \frac{n(n-3)}{2}
\end{align} 
For any set of particle labels $I \subset \{1,2, \ldots n\}$ one can define Mandelstam variables as follows,
\begin{align}
s_I = \bigg( \sum_{i \in I} p_i \bigg)^2 = \sum_{i, j \in I; \ i < j} s_{ij} 
\end{align}

\begin{figure}[h]
\centering 
\includegraphics[width=4cm]{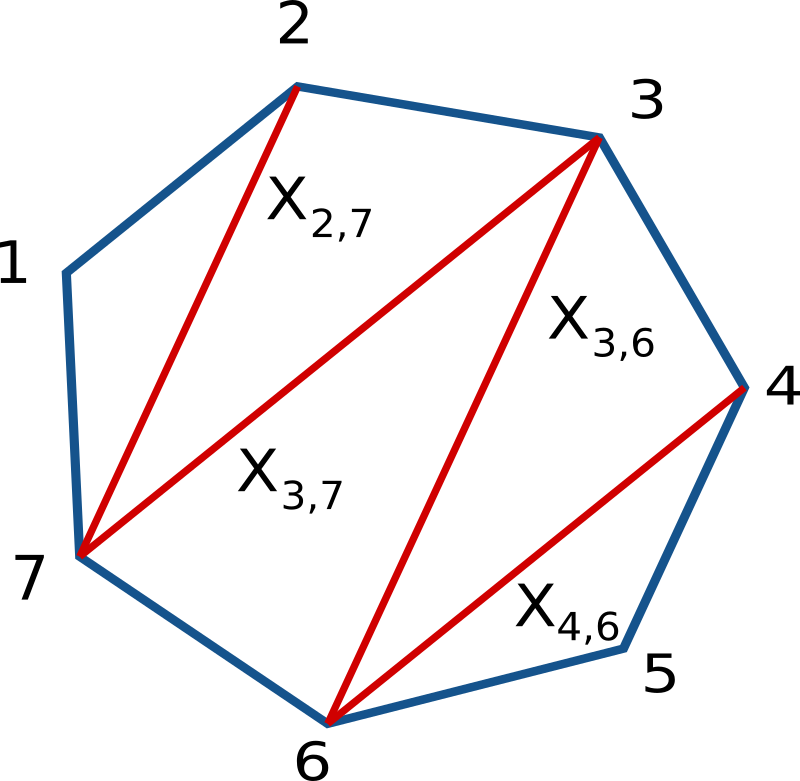} 
\caption{\label{fig: planar-X} Planar variables. }
\end{figure}

\subsection{Planar kinematic variables and the scattering form}
For cyclically ordered particles it's useful to define planar kinematic variables,
\begin{align}
X_{i, j} = s_{i, i+1, \ldots j-1} ; \quad 1\le i < j \le n .
\end{align} 
From the definition it is easy to see that $X_{i, i+1} = 0$ and $X_{1, n} =0$. These variables  $X_{i, j}$ can be visualized as diagonal between $i^{th}$ and $j^{th}$ vertices of the corresponding $n$-gon (see figure \ref{fig: planar-X}).

These variables are related to Mandelstam variables via following relation.
\begin{align}
s_{ij} = X_{i, j+1} +  X_{i+1, j} -  X_{i, j} -  X_{i+1, j+1}
\end{align}

\begin{figure}[H]
	\centering 
	\includegraphics[width=9cm]{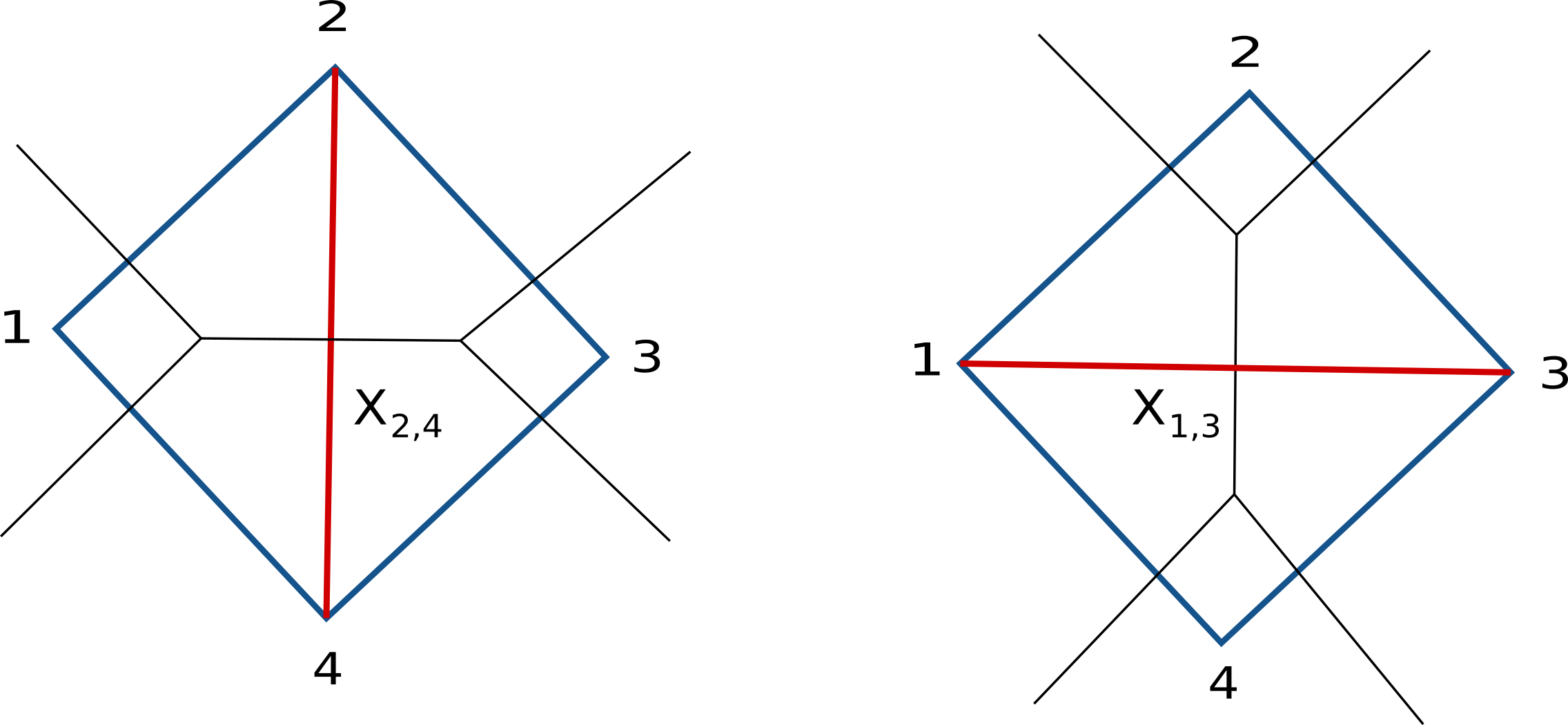} 
	\caption{\label{fig: s-t_channels} A planar variable cuts an  internal propagator of the Feynman diagram once. }
\end{figure}

In other words $X_{i, j}$  are dual to $\frac{n(n-3)}{2}$ diagonals of $n$-gon made up of  edges with momenta $p_1, p_2, \ldots p_n$. Each diagonal \emph{i.e}  $X_{i, j}$ cuts the internal propagator of a Feynman diagram once (see figure \ref{fig: s-t_channels}). Thus there exists an one-to-one correspondence between cuts of cubic graphs and complete triangulations of a $n$-gon.
\vspace{0.5 cm}
A partial triangulation of regular $n$-gon is a set of non-crossing diagonals which do not divide the $n$-gon into $(n-2)$ triangles. Here is an example of partial triangulation for a $5$-gon. \\

\begin{figure}[H]
	\centering 
	\includegraphics[width=10 cm]{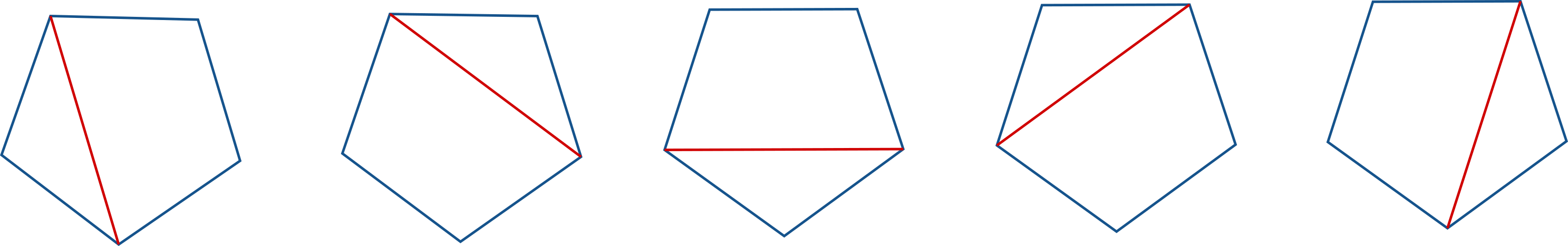} 
	\caption{\label{fig: partial-triang} Partial triangulations of a pentagon. }
\end{figure}

\begin{figure}[H]
	\centering 
	\includegraphics[width=7cm]{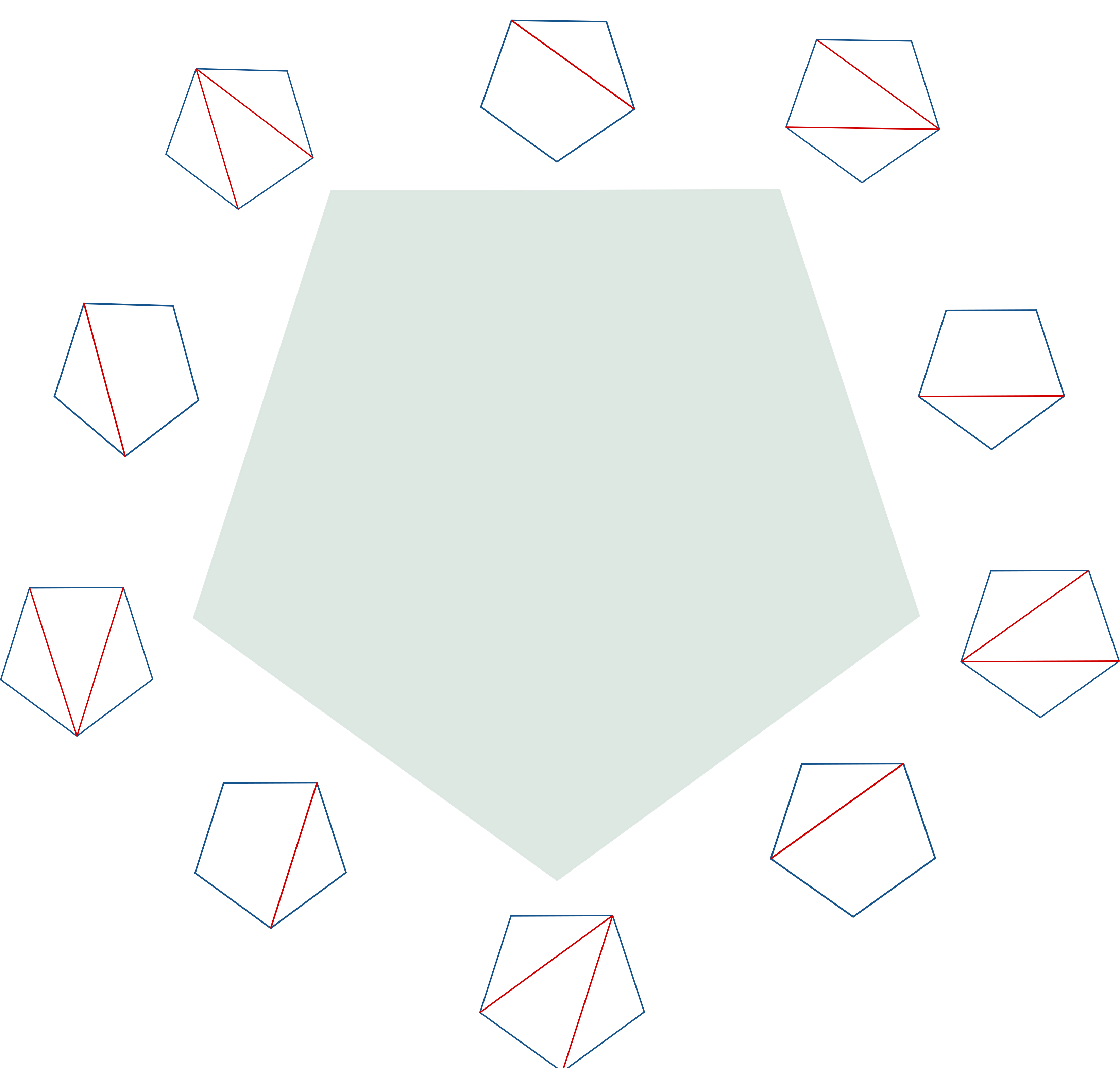} 
	\caption{\label{fig: pentagon} Two dimensional associahedron $\mathcal{A}_5$ :  5 partial triangulations are represented by 5 diagonals. 5 complete triangulations are represented by 5 vertices. }
\end{figure}

The \emph{associahedron} of dimension $(n-3)$ is a polytope whose co-dimension $d$ boundaries are in one-to-one correspondence with the partial triangulation by $d$ diagonals (see figure \ref{fig: pentagon}). 
\vspace*{0.5cm}
The vertices  represent complete triangulations and $k$-faces represent $k$-partial triangulations of the $n$-gon. The total number of ways to triangulate a convex $n$-gon by non-intersecting diagonals is the $(n - 2)$-th Catalan number, $C_{n-2} = \frac{1}{n-1} \binom {2n-4}{n-2}$, a solution found by Euler. The dimension of the associahedron corresponding to a $n$-gon is $(n-3)$. 

Now we introduce the planar scattering form, a differential form on the space of kinematic variables $X_{i,j}$ that encodes  information about on-shell tree-level scattering amplitudes of the scalar $\phi^3$ theory. Let $g$ denote a (tree) cubic graph with propagators $X_{i_a,j_a}$ for $a=1,\ldots, n{-}3$. The ordering is important here. For each ordering of these propagators, one assigns a value $\text{sign}(g)\in \{\pm 1\}$ to the graph with the property that flipping two propagators flips the sign. The form must have logarithmic singularities at $X_{i_a,j_a} =0$. Therefore one assigns to the graph a $d\log$ form and thus defines the {\it planar scattering form} of rank $(n{-}3)$ :
\begin{equation}\label{eq:planar_form}
\Omega^{(n{-}3)}_n:= \sum_{{\text{planar } g} } \text{sign}(g) \bigwedge_{a=1}^{n{-}3} d\log X_{i_a,j_a}
\end{equation}
where the sum is over each planar cubic graph $g$. It's important to note that there are two sign choices\footnote{For `clockwise' or `anticlockwise' ordering of propagators $g = +1$ or $-1$, respectively.} for each graph.  Due to this fact there are many different scattering forms. But one can fix the scattering form uniquely\footnote{Actually the requirement of projectivity fixes the scattering form up to an overall sign which one ignores.} if one demands projectivity of the differential form \emph{i.e.} if one requires the form should invariant under {\it local} $GL(1)$ transformations $X_{i,j}\rightarrow \Lambda(X) X_{i,j}$ for any index pair $(i,j)$. We use this projectivity property to define a useful operation called \emph{mutation}.
\begin{figure}[h]
	\centering 
	\includegraphics[width=11cm]{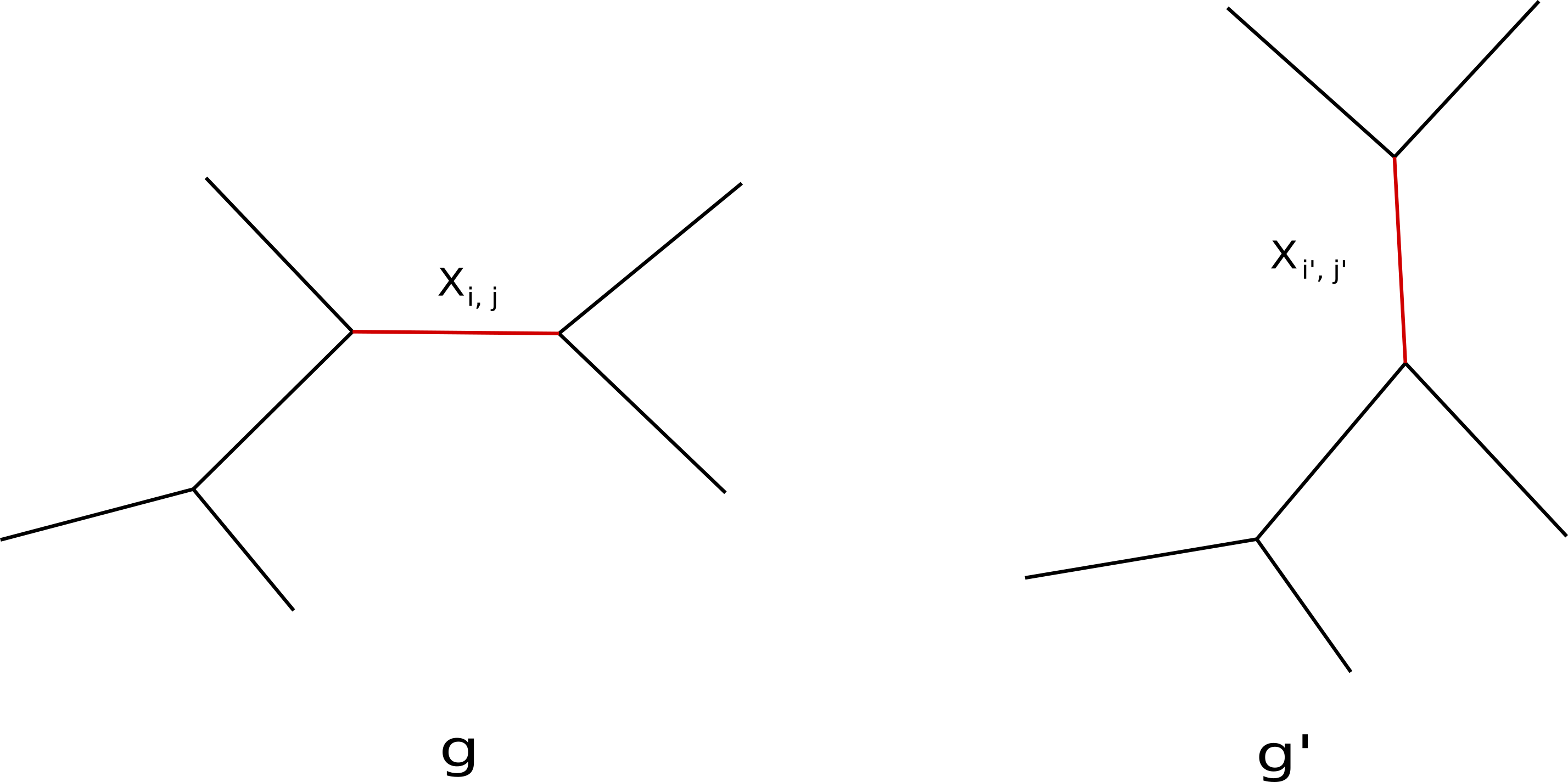} 
	\caption{\label{fig: mutation} Two 5-point graphs related by mutation : $X_{i,j} \to X_{i',j'}$. }
\end{figure}
 Two planar graphs $g$ and $g'$ are related by a \emph{mutation} if we can obtain one from the other just by exchanging four-point sub-graph channel (see figure \ref{fig: mutation}). In that figure \ref{fig: mutation}, $X_{i,j}$ and $X_{i',j'}$  are the mutated propagators of the graphs $g$ and $g'$, respectively. Let's denote the rest of the (common) propagators as $X_{i_b,j_b}$ with $b = 1, 2, \ldots n-4$. Under a local GL(1) transformation, the $\Lambda(x)$ dependence of the scattering form becomes,
\begin{equation}
 \big(\text{sign}(g) +  \text{sign}(g')\big) \, d\log \Lambda \, \wedge \bigwedge_{a=1}^{n{-}4} d\log X_{i_a,j_a} + \ldots
\end{equation}
But since we demand projectivity the form shouldn't have any $\Lambda(x)$ dependent piece and therefore,
\begin{equation}
\text{sign}(g') = - \,  \text{sign}(g)
\end{equation}
Note that projectivity ensures that the form should be ratios of Mandelstam variables. Here are few examples of $(n-3)$-forms in kinematic space of $n$ particle scattering.
\begin{align}
\label{eq:planar_form_4}
\Omega^{(1)}_{n =4}
 = d\log\left(\frac{s}{t}\right)=d\log \left(\frac{X_{1,3}}{X_{2,4}}\right)
\end{align}
\begin{align}\label{eq:planar_form_5}
\Omega^{(2)}_{n = 5}
=&~~~d\log \frac{X_{1,3}}{X_{2,4}} \wedge d\log \frac{X_{1,3}}{X_{1,4}} + d\log \frac{X_{1,3}}{X_{2,5}} \wedge d\log \frac{X_{3,5}}{X_{2,4}}\,
\end{align}
and so on. 

\subsection{The kinematic associahedron}
Above we described how one gets an associahedron $\mathcal{A}_n$ in the kinematic space $\mathcal{K}_n$, but it is not evident how it should be embedded  in $\mathcal{K}_n$. Because $\mathcal{K}_n$ and $\mathcal{A}_n$ are of different dimensionality
\begin{align}
dim (\mathcal{K}_n) = \frac{n(n-3)}{2} \\
dim (\mathcal{A}_n) = n-3
\end{align}

One needs to impose constraints to embed $\mathcal{A}_n$ inside $\mathcal{K}_n$. One natural choice is to demand all planar kinematic variables to be positive,
 \begin{align}
 X_{i,j} \ge 0 \ ; \quad 1\le i < j \le n
 \end{align}
These are $\frac{n(n-3)}{2}$ inequalities and thus cutout a big simplex ($\Delta_n$) inside $\mathcal{K}_n$ which is still $\frac{n(n-3)}{2}$ dimensional. Therefore one needs  $\frac{n(n-3)}{2} -(n-3) = \frac{(n-2)(n-3)}{2}$ more constraints to embed the $\mathcal{A}_n$ inside $\mathcal{K}_n$. To do that one imposes the following constraints,
 \begin{align}
s_{ij} = - \, c_{ij} \ ;  \quad  for \ \ 1\le i < j \le n-1, \ |i-j|\geq 2
\end{align}
where $c_{ij}$ are positive constants. 

These constraints give a space  $H_n$ of dimensions $(n-3)$ which is precisely the dimension of $\mathcal{A}_n$. The kinematic associahedron $\mathcal{A}_n$ now can be embedded in $\mathcal{K}_n$ as the intersection of the simplex $\Delta_n$ and the subspace $H_n$ as follows,
 \begin{align}
 \mathcal{A}_n := H_n \cap \Delta_n 
\end{align}

Once one has the associahedron in $\mathcal{K}_n$ all one needs to do is to obtain its canonical form $\Omega(\mathcal{A}_n)$. Since associahedron is a simple\footnote{A polytope $\mathcal{A}_n$ is called \emph{simple } if each of its vertex is adjacent to $d$ facets where $d= dim(\mathcal{A}_n)$. Its easy to see associahedron satisfies the criterion and hence is an example of simple polytope.} polytope  one can directly write down its canonical form as follows \cite{Arkani-Hamed:2017tmz}.
\begin{equation}\label{eq:simple-form}
\Omega(\mathcal{A}_n)=\sum_{\text{vertex }Z} \text{sign}(Z)\bigwedge_{a=1}^{n{-}3}d\log X_{i_a,j_a}
\end{equation}
where  for each vertex $Z$ and  $X_{i_a,j_a}= 0$ denote its adjacent facets\footnote{One should be careful about the orientations of the facets. Depending on the ordering of the facets are assigned a sign$(Z) \in \{\pm 1\}$.} for $a=1,\ldots, n{-}3$. The claim is the above differential form \eqref{eq:simple-form} is identical to the pullback of scattering form \eqref{eq:planar_form} (in $\mathcal{K}_n$) to the subspace $\mathcal{H}_n$. We can justify this statement by identifying  : $g \leftrightarrow Z$ and sign$(g)$ $\leftrightarrow$ sign$(Z)$. 

\begin{itemize}
\item There is a one-to-one correspondence between vertices $Z$ and planar cubic graphs $g$. Also $g$ and its corresponding vertex $Z$ has same propagators $X_{i_a,j_a}$.

\item  Let $Z$ and $Z'$ be two vertices related by mutation. Note that mutation can also be framed in the language of triangulation.  Two triangulations are related by a mutation if one can be obtained from the other by exchanging exactly one diagonal (see figure \ref{fig: planar-mutation}). 

\begin{figure}[h]
	\centering 
	\includegraphics[width=9cm]{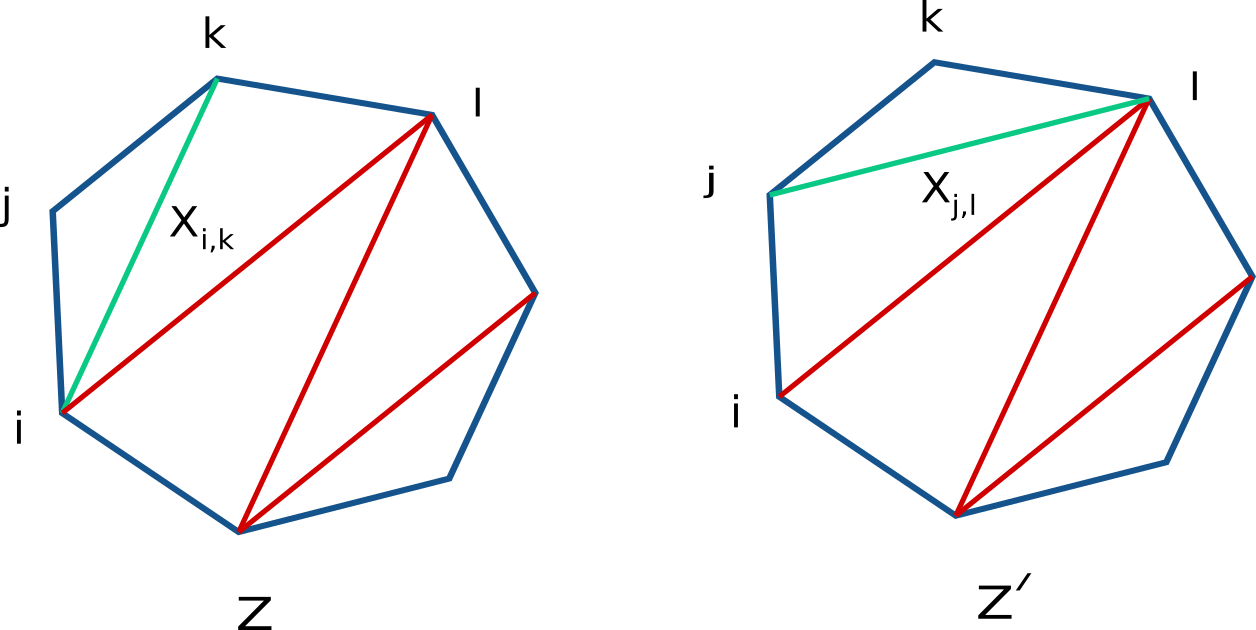} 
	\caption{\label{fig: planar-mutation} Two triangulations related by mutation : $X_{i,k} \to X_{j,l}$. }
\end{figure}

Thus for $Z$ and $Z'$ vertices we have 
\begin{align}\label{eq:mutation-triang}
\bigwedge_{a=1}^{n{-}3}dX_{i_a,j_a} = - \bigwedge_{a=1}^{n{-}3}dX_{i_a',j_a'}
\end{align}
which leads to sign-flip rule identical to $g$ \emph{i.e.}  sign$(Z)$ $= -$  sign$(Z')$.
\end{itemize}
Therefore one can construct the following quantity (an $(n-3)$-form) which is independent of $g$ on pullback.
\begin{equation}\label{eq:dX}
d^{n{-}3}X:=\text{sign}(g)\bigwedge_{a=1}^{n{-}3}d X_{i_a,j_a}
\end{equation}
Substituting this in \eqref{eq:simple-form} one gets,
\begin{equation}\label{eq:amp_form}
\Omega(\mathcal{A}_n)= \underbrace{\left(\sum_{\text{planar }g}\frac{1}{\prod_{a=1}^{n{-}3}X_{i_a,j_a}}\right)}_{\mathcal{M}_n}d^{n{-}3}X
\end{equation}
where $\mathcal{M}_n$ is the expected tree level planar $n$-point scattering amplitude for scalar cubic theory.

\section{Positive geometry for $\phi^{4}$ interactions}

As reviewed in the previous section, the relationship between (planar) Feynman graphs in $\phi^{3}$ theory and positive geometry (namely associahedron) encapsulates a few intriguing facts.
 
\noindent {\bf (1)} There is a one to one correspondence between Feynman graphs with complete triangulations of a polygon. 

\noindent {\bf (2)} Dimension of the kinematic associahedron is the same as number of propagators in an $n$-particle scattering. 

\noindent {\bf (3)} Each co-dimension $k$ facet of the associahedron is in one to one correspondence with a $(n-3-k)$-partial triangulation of the $n$ sided polygon. 

At first sight, it is tempting to consider a generalisation of these inter-relationships between polygons and planar (tree-level) amplitudes in $\phi^{4}$ theory. 

\begin{figure}[h]
 \centering
\includegraphics[width= 10cm]{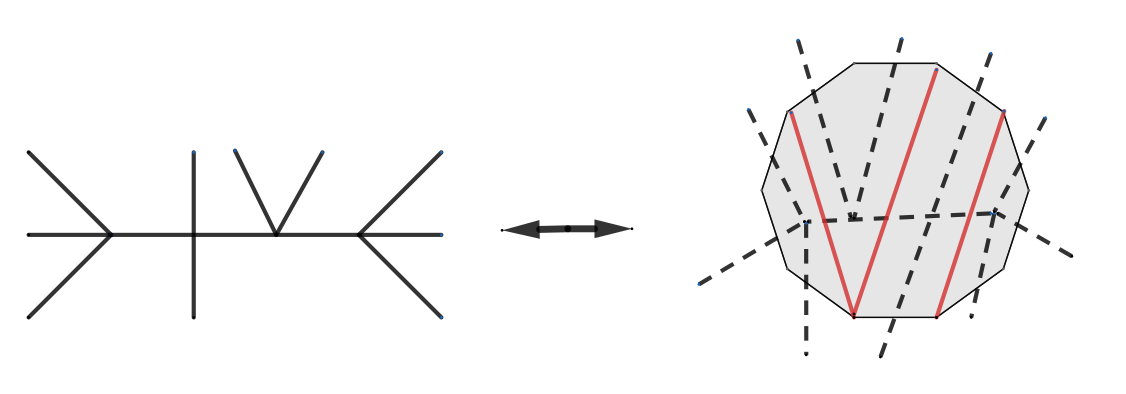} 
\caption{\label{fig: qcuts} A one-to-one correspondence between Feynman graphs of $\phi^{4}$ theory and quadrangulations of an even polygon.}
\end{figure}

One immediately notices the following. Precisely as in the case of $\phi^{3}$ theory and the triangulations of polygon, there is a one-to-one correspondence between planar tree-level diagrams of $\phi^{4}$ theory and  \emph{complete quadrangulations}\footnote{By complete quadrangulation we just means decomposing a polygon into maximum number of quadrilaterals. We will refer to any subset of the diagonals which do not constitute a complete quadrangulation as partial quadrangulation.} of a polygon (see figure \ref{fig: qcuts}).
\vspace{-0. cm}

\begin{figure} 
	 \centering
\includegraphics[width=11cm]{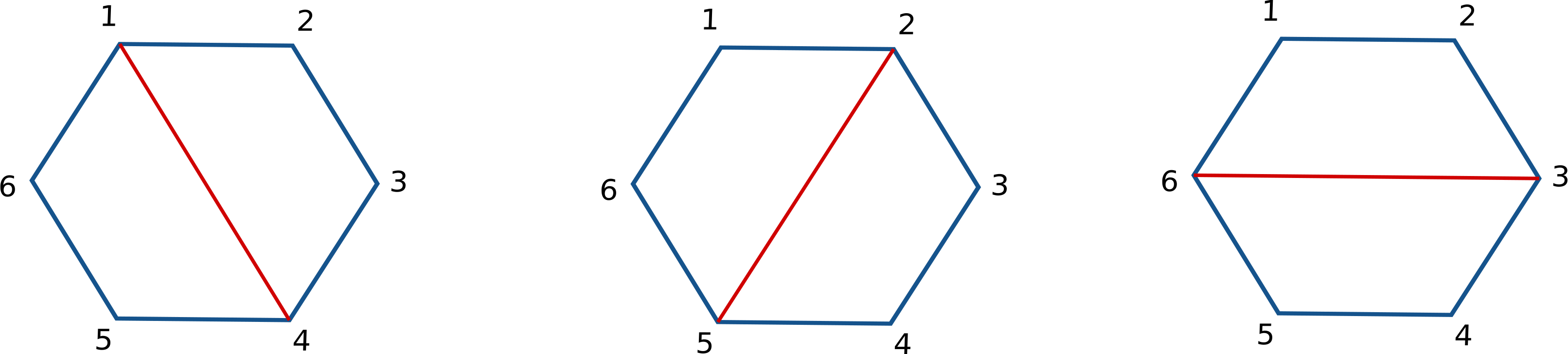} 
\caption{\label{fig: 3channels} The 3 different planar channels for 6-point scattering.}
\end{figure}

A few facts about the quadrangulations are well known \cite{chapoton}.
The total number of quadrangulations of an $n\, =\, ( 2N+2)$-gon is given by the \emph{Fuss-Catalan} number,
\begin{center}
$F_{N} = \frac{1}{2N+1}  {3 N \atop } C_{N} .$
\end{center}
We can thus ask the following question. Is there a polytope ${\cal S}_{n}$ whose vertices are in $1-1$ correspondence with all quadrangulations of a polygon and whose dimension is same as the number of propagators in a single channel as in the associahedron case. Since, each quartic graph with $n\ =\ 2N +2$ external legs has precisely $N-1$ propagators, 
\begin{align*}
dim({\cal S}_{n})= N-1 = \frac{n-4}{2}.
\end{align*}
We can now ask if there is a polytope whose dimension is $\frac{n-4}{2}$ and number of vertices are same as $F_{N}$. 
Here we immediately run into an obstacle due to the fact that for the six-point scattering (\emph{i.e.} $N=2$) we should get a one dimensional polytope, which can only be a line segment with two boundaries but since there are in fact \emph{three} planar scattering channels (see figure \ref{fig: 3channels}) for the six-point diagram we cannot find such a polytope with boundaries which correspond to all three propagators going onshell.
So, the only way to define a polytope is to exclude one of the channels using some systematic rule. This idea was precisely encapsulated in \cite{Baryshnikov} in a different context and used to construct the Stokes polytope.
\subsection{Stokes polytope}\label{q-comp}
In order to introduce Stokes polytope, we first need to define a notion of $Q$-compatibility which selects, among the set of all (complete) quadrangulations of a polygon, a subset which will be in one-to-one correspondence with vertices of Stokes polytope.

\begin{figure}[H]
\centering 
\includegraphics[width=12cm]{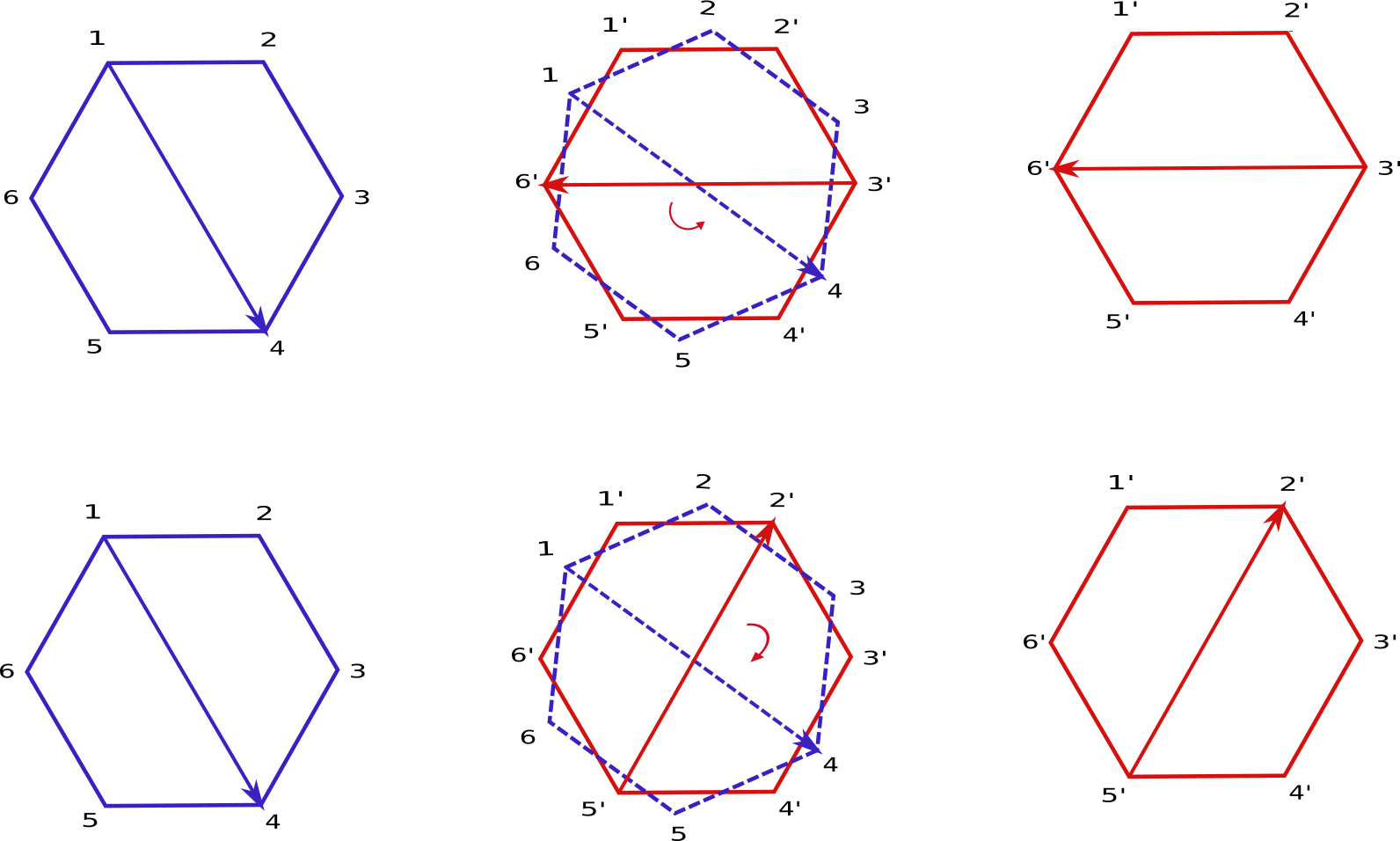} 
\caption{\label{fig: Q-compatability} The above figure shows 36 is $Q$-compatible with 14 but 25 is not. }
\end{figure}

Consider, a pair of quadrangulations $Q$ and $Q'$ of a regular 2$N$+2 gon which we call \emph{blue} and \emph{red} respectively with diagonals directed from odd to even vertices (see figure \ref{fig: Q-compatability}). We want to define a rule to check if $Q'$ is compatible to a given $Q$. Here is the rule :  We first rotate $Q$ (blue) anti-clockwise and then superimpose it over $Q$ so that the vertices now get interlaced.
We then say $Q'$ is Q-compatible with $Q$ if and only if at each crossing of diagonals the pair (red,blue) in that order are oriented anti-clockwise.

We must emphasise that $Q$-compatability is \emph{not} an equivalence relation and is very much dependent on the reference quadrangulation $Q$, as can be easily checked that 14 is compatible with 36, 25 with 14 and 36 with 25 \footnote{ A simple way to remember this rule is that every diagonal is Q-compatible with every alternate diagonal when we move clockwise(14 with 36 , 25 with 41 and 36 with 52).}.\\
\begin{figure}[H]
	\centering 
	\includegraphics[width=12cm]{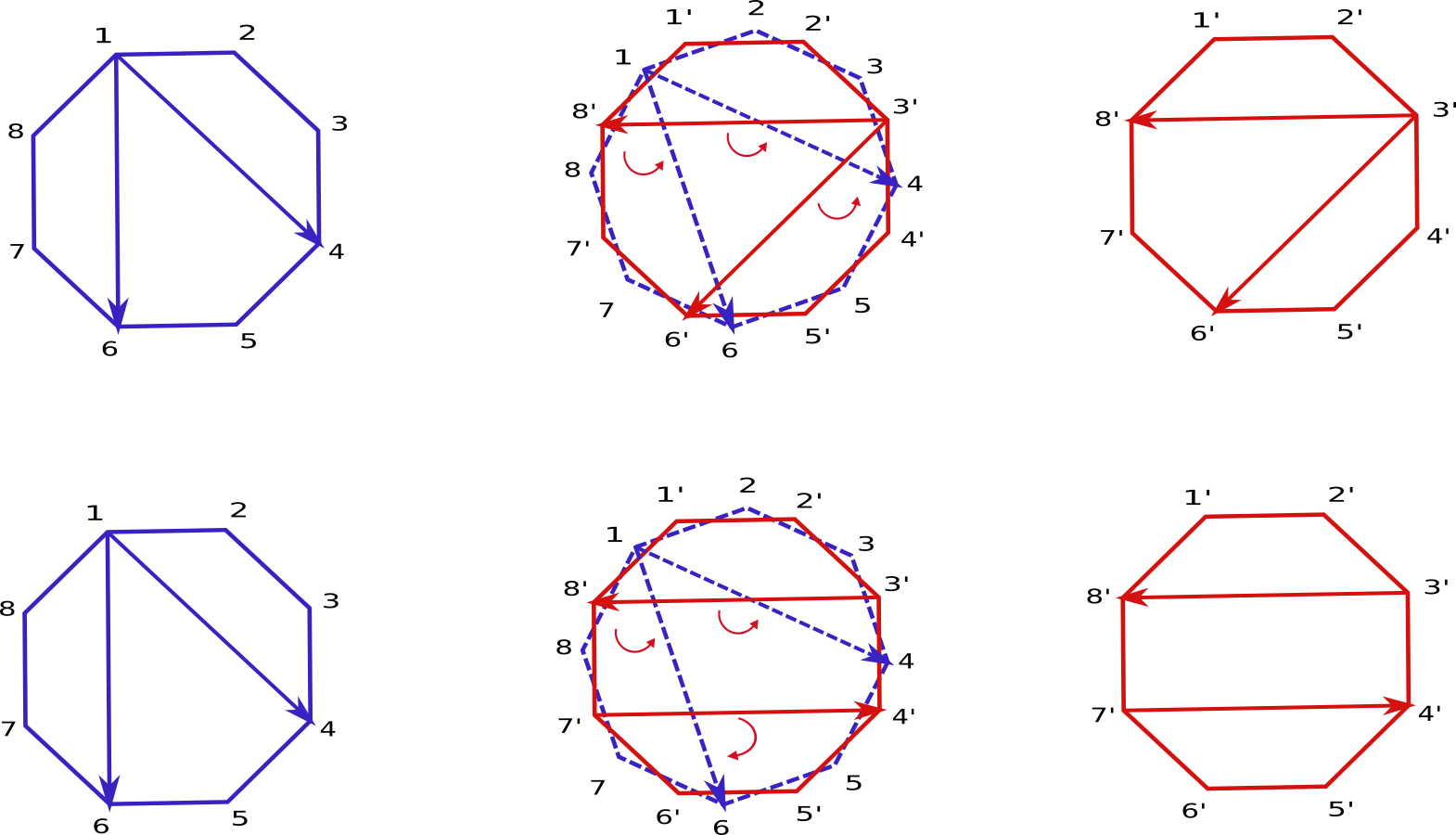} 
	\caption{\label{fig: Q-compatability2} The above figure shows $\{36,38\}$ is $Q$-compatible with $\{14,16\}$  but $\{38,47\}$ is not. }
\end{figure}

We can now define a {\it flip} as the replacement of a diagonal of any hexagon inside the quadrangulation of the polygon with its Q-compatible diagonal, this corresponds to changing to a compatible channel for any $6$-point diagram inside our $(2 N+2)$-point diagram. This is the analogue of mutation for quartic case (see eqn. \eqref{eq:mutation-triang}).

We can now define the Stokes polytope ${\cal S}_{n}^{Q}$ simply by starting with a particular quadrangulation $Q$ with diagonals $(i_1 j_1,..., i_{N-1} j_{N-1}),\; N \ge 3$ and by performing flips on each diagoanl $i_k j_k$ sitting inside the hexagon with vertices $\{i_{k-1}, i_{k}, i_{k+1}, j_{k+1}, j_{k}, j_{k-1} \}$ iteratively till we do not generate any new quadrangulations. We illustrate this for the $N=3$ (8-point scattering) below.
\begin{figure}[H]
	\centering
\includegraphics[width=18cm]{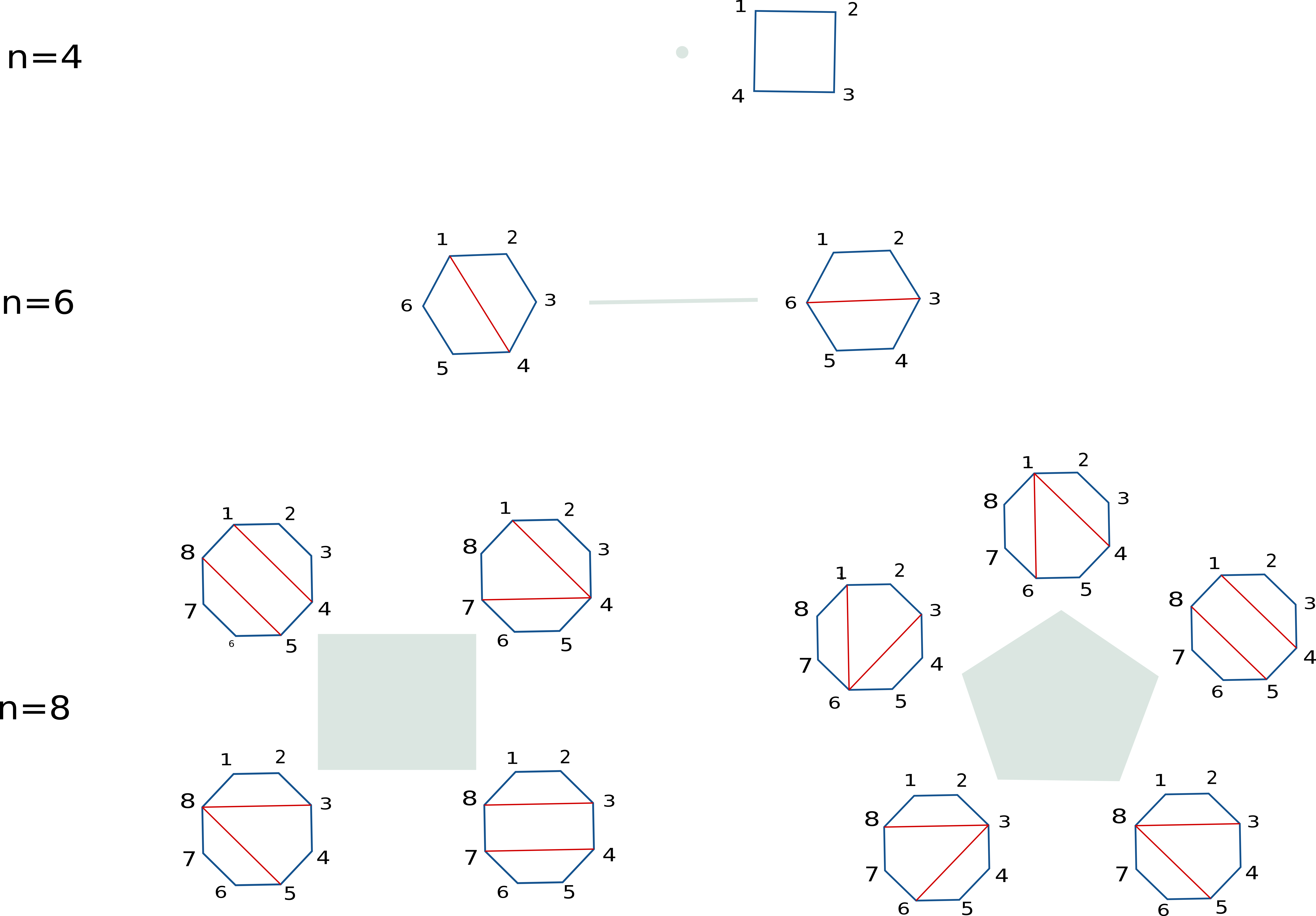} 
\caption{\label{Stokes-poly} The first few Stokes polytopes. Note that for $n=8$ there are two kinds of  polytopes. This is one of the key features of the quartic case.}
\end{figure}
\vspace{0.5 cm}
We start with the $Q =\{14, 58 \}$ and flip either $14$ to $38$ in $\{1,2,3,4,5,8\}$ or $58$ to $47$ in $\{1,4,5,6,7,8\}$ and to get $Q_1 =\{36, 58 \}$ or  $Q_2 =\{14, 47 \}$ respectively, then a further flip of either $14$ to $38$ in $\{1,2,3,4,7,8\}$ or $58$ to $47$ in $\{3,4,5,6,7,8\}$ both give $Q_4 = \{16, 47 \}$. Further flips do not give us any new  quadrangulations. Thus the corresponding Stokes Polytope in this case has 4 vertices. This is shown in the left half of $n=8$ in figure \ref{Stokes-poly}.

If we start with $Q =\{14, 16 \}$ and flip either $14$ to $36$ in $\{1,2,3,4,5,6\}$ or $16$ to $58$ in $\{1,4,5,6,7,8\}$ to get $Q_1 =\{36, 16 \}$ or  $Q_2 =\{14, 58 \}$ respectively, then  further flips of $16$ to $38$ in $\{1,2,3,6,7,8 \}$ and $14$ to $38$ in $\{1,2,3,4,5,8 \}$ give $Q_4 = \{36, 38\}$ and $Q_5 = \{36, 58\}$. Further flips do not give us any new compatible quadrangulations\footnote{Notice that if one flips 58 keeping 38 fixed in $Q_5 = \{36, 58\}$, one gets $\{38, 47\}$. But it's \emph{not} compatible with $Q =\{14, 16 \}$ (see fig. \ref{fig: Q-compatability2}).}.  Thus the corresponding Stokes polytope in this case has 5 vertices. This is shown in the right half of $n=8$ in figure \ref{Stokes-poly}.
\paragraph{{\it It can be checked that if we start with any of the $F_3 =12 $ quadrangulations then the Stokes polytope we get is either a square or a pentagon. This is easily seen if we notice that the other $10$ quadrangulations can be obtained from $\{14, 16 \}$ and $\{14, 58 \}$ by cyclic permutations and thus just amount to relabeling of the vertices.}}\label{8primitive}

We can proceed along these lines to obtain Stokes polytopes for any $n= 2N+2$, and there will be several Stokes polytopes depending on the reference quadrangulation $Q$ we start with. Some of them do turn out be associahedra and we will say more about this in  appendix \ref{appendixA}.
 We can thus sumarize the Stokes polytope in analogy with associahedron as follows: \\
\begin{center}
  {\bf Vertices} $\leftrightarrow$ {\bf Q-compatible quadrangulations }\\
  {\bf Edges} $\leftrightarrow $ {\bf Flips between them } \\
  {\bf k-Facets} $\leftrightarrow $ {\bf k-partial quadrangulations} 
\end{center}

As we see, there are two key differences in the relationship of the Stokes polytope with quadrangulations from that of the associahedron and triangulations. First being, definition of Stokes polytope depends on the reference quadrangulation $Q$, and for each $Q$ one has a  Stokes polytope ${\cal S}_{n}^{Q}$. Secondly  vertices of ${\cal S}_{n}^{Q}$ are not in 1-1 correspondence with all the quadrangulations of the polygon but only with a specific sub-set of them, namely $Q$-compatible quadrangulations. As all (planar) diagrams of a $\phi^{4}$ theory are in 1-1 correspondence with set of \emph{all} quadrangulations of a polygon, it is clear that a single ${\cal S}_{n}^{Q}$ can not be the amplituhedron for planar $\phi^{4}$ theory. 

However a rather enticing feature of definition of ${\cal S}_{n}^{Q}$ is a notion of the flip, which is analogous to mutation in the case of triangulations. As it was the mutation which was responsible for defining a unique scattering form in ${\cal K}_{n}$ in the $\phi^{3}$ case, there is a possibility that the flip may do the same in this case. In the next section we propose just such a definition of planar scattering form for $\phi^{4}$ theory in kinematic space, which however will depend on the reference quadrangulation $Q$. 

\section{Planar scattering form for $\phi^{4}$ interactions}

We consider tree level scattering amplitudes in a massless scalar field theory with quartic interactions.  Given a specific ordering of external particles, we consider contribution of only planar diagrams which are consistent with this ordering.\footnote{By tree-level planar diagrams we mean diagrams with no crossing.} We refer to such amplitudes as \emph{planar amplitudes} of massless $\phi^{4}$ theory.  These amplitudes can be thought of as analogs of the partial amplitudes ${\cal M}_{n}(\alpha\vert \alpha)$ in the context of bi-adjoint scalar $\phi^{3}$ theory\footnote{It is conceivable that the amplitudes we analyse can be considered as basic building blocks of amplitudes of a bi-adjoint scalar field theory with quartic interaction of the type $\textrm{Tr}\left[ [\phi, \phi]^{2}\right]$ where $[\phi,\phi]$ is the bi-adoijnt Lie bracket given by $f^{ijk}\tilde{f}^{i^{\prime}j^{\prime}k^{\prime}}\phi^{i i^{\prime}}\phi^{j j^{\prime}}$. However as bi-adjoint scalar theory with quartic interaction has not been considered in literature so far, we will not refrain from exploring this point of view further.}  which was considered in \cite{Arkani-Hamed:2017mur}. 

We would like to extend the idea of defining planar scattering form to planar amplitudes in massless $\phi^{4}$ theory.
However a quick look at the simplest example of six point amplitude shows us that such a form can not be projective.
In general, for an $n$ particle amplitude in quartic theory, the number of planar diagrams can be even or odd and there is no sense in which projectivity can be employed to fix a unique scattering form.  
In the absence of projectivity, it is a priori not clear how do we define a planar scattering form for planar amplitudes in $\phi^{4}$ theory. The hint in our case (that we alluded to in the previous section) comes from one of the key observations made in \cite{Arkani-Hamed:2017mur}. Namely,  defining a scattering form projectively is equivalent to choosing the relative signs among various terms via mutation, which is in turn equivalent to flipping one of the diagonals in the triangulation of the $n$-gon. 

For $\phi^{4}$ interaction,  even though mutation or projectivity do not appear to be relevant concepts, as we saw above,  there is an analog. Given a reference quadrangulation $Q$, there is a set $Q$-compatible quadrangulations for which a notion of flip is well defined. Whence given a $Q$ and its corresponding set of $Q$-compatible quadrangulations, we can define a planar scattering form on the kinematic space ${\cal K}_{n}$ as follows. 

Let $Q$ be a quadrangulation of an $n$-gon which is associated to an planar Feynmann diagram with propagators given by $X_{1},\dots, X_{\frac{n-4}{2}}$. Then we define the ($Q$-dependent) planar scattering form as, 
\begin{equation}\label{psf}
\begin{array}{lll}
\Omega^{Q}_{n}\ = \displaystyle \, \sum_{\textrm{flips}} (-1)^{\sigma(\textrm{flip})}  d \ln X_{i_{1}}\wedge\dots d \ln X_{i_{\frac{n-4}{2}}} 
\end{array}
\end{equation}
where $\sigma(\textrm{flip})\ =\, \pm 1$ depending on whether the quadrangulation $X_{i_{1}},\dots, X_{i_{\frac{n-4}{2}}}$ can be obtained from $Q$ by even or odd number of flips. 

As the set of $Q$-compatible quadrangulations (for a given $Q$) does not exhaust all quadrangulations or equivalently,  all the planar Feynman diagrams,  the set of terms which appear in the planar scattering form in eqn. \eqref{psf} does not correspond to all the diagrams of the theory. As an example consider $N=6$ case and let $Q\ =\ 14$. Then the set of $Q$ compatible quadrangulations are $\{\,(14\ , +),\ (36\ , -) \}$. We have attached a sign to each of the quadrangulation which measures the number of flips needed to reach it starting from reference $Q\ =\ 14$.
 Whence the form $\Omega^{Q}_{6}$ on the kinematic space is given by, 
\begin{equation}\label{1formfor6}
\Omega^{Q=(14)}_{6}\ =\  ( d\ln X_{14}\ -\ d\ln X_{36})
\end{equation}

It is clear that this form does not capture singularity associated to $X_{25}$ channel for the 6 particle amplitude. Hence it may appear that eventually we may not recover full planar scattering amplitude from such a form. However there are two more $Q$s we need to consider. For $Q\ =\ 36$ the $Q$-compatible set is $\{ (36, +), (25, -)\}$ and for $Q\ =\ 25$ the $Q$-compatible set is $\{(25,+), (14,-)\}$. The corresponding forms on Kinematic space are given by

\begin{equation}\label{1formfor6'}
\begin{array}{lll}
\Omega^{Q=(36)}_{6}\ =\ ( d\ln X_{36}\ -\ d\ln X_{25})\\
\Omega^{Q=(25)}_{6}\ =\  ( d\ln X_{25}\ -\ d\ln X_{14})
\end{array}
\end{equation}

Hence we see that unlike the planar scattering form in the case of $\phi^{3}$ interaction which is uniquely determined by requirement of projectivity, we have $F_N$ planar scattering forms, one for each quadrangulation.

It can be easily checked that  for all $Q$, $\Omega^{Q}_{n}$ in eqn. \eqref{psf} factorizes correctly when any one of the channels goes on-shell. For $ i <  j$, 
\begin{equation}
\Omega^{Q}_{n}\bigg|_{X_{ij}\ \rightarrow\ 0}\ =\ \Omega^{Q_{1}}_{\vert j-i+1\vert}(i,i+1,\dots,j)\ \wedge\ \frac{d X_{ij}}{X_{ij}}\ \wedge\ \Omega^{Q_{2}}_{n+2-\vert j - i +1\vert}(j,\dots,n,1,\dots, i)
\end{equation}
with $Q_{1},\ Q_{2}$ are quadrangulations associated to the polygons $\{(i,i+1,\dots,j),\ (j,\dots,n,1,\dots, i)\}$ respectively.

A happy fact about $\Omega^{Q}_{n}$ will emerge in  the next section : Paralleling the construction of \cite{Arkani-Hamed:2017mur} we will see how these forms naturally descends to the  canonical form on a ${\cal S}_{n}^{Q}$ : As Stokes polytope is a positive geometry, it has a canonical form associated to it which has  (logarithmic) singularities on all the facets, such that the residue of restriction of this form on any of the facet equals the canonical form on the facet. (see appendix in \cite{Arkani-Hamed:2017mur} and \cite{Arkani-Hamed:2017tmz} for details regarding canonical form on positive geometries.)  

Stokes polytopes are simple\footnote{The way Stokes polytopes are defined they are always \emph{simple}. The reason is the following. Any vertex of the polytope represents a complete quardangulation. The number of diagonals needed to complete the quardangulation of an $n$-gon is $\frac{n-4}{2}$. This is also the number of dimensions of the corresponding Stokes polytope. Now to get the facets (co-dimension one boundaries) one needs to remove one of those $\frac{n-4}{2}$ diagonals, which can be done in exactly $\frac{n-4}{2}$ different ways. Thus the number of facets attached to a given vertex of Stokes polytope matches its dimension.} polytopes. But an explicit formula for canonical form on ${\cal S}_{n}^{Q}$ does not seem to be available in the literature. The planar scattering form defined above however gives us precisely such a form on ${\cal S}_{n}^{Q}$.  That is, we will take a cue from ideas of \cite{Arkani-Hamed:2017mur} and start with a definition of planar scattering form for $\phi^{4}$ theory and show that it descends to a form on ${\cal S}_{n}^{Q}$ which satisfies all the properties required of the canonical form.

\section{Locating the Stokes polytope in kinematic space}\label{locat=n6section}

In this section we realise Stokes polytopes $\{{\cal S}_{6}^{Q}\ \vert\ Q\ \in (14,25,36)\}$ for 6 particle amplitude as positive geometries in kinematic space. We show how the planar scattering form $\Omega_{n}^{Q}$ defined above descends to the canonical form on ${\cal S}_{6}^{Q}$.
Before proceeding we once again emphasise that, there are several Convex realisations of Stokes polytopes. Their realisation as a simple polytope is given in \cite{Baryshnikov, chapoton}, as well as in a beautiful recent work \cite{Palu}. Although we consider convex realisations of only 2 and 3 dimensional Stokes polytopes, such convex realisation exists for all $n$  as shown in \cite{Palu}.
More in detail, we  explicitly study  convex realisations of lower dimensional Stokes polytopes for $n=6, 8$ and $10$ respectively. Our strategy is to embed the Stokes polytopes ($S^Q_n$)  inside corresponding associahedra ($\mathcal{A}_n$)  for given number of particle $n$. A more precise formulations of our idea which appears to generalise our construction for arbitrary $n$ has appeared recently in mathematics literature \cite{Palu}

We proceed exactly as in \cite{chapoton, Baryshnikov}. That is we begin by fixing a reference quadrangulation $Q$  in terms of kinematic data (\emph{i.e.} a set of $X_{ij}'s$) and get  a Stokes polytope ${\cal S}_{n}^{Q}$  in ${\cal K}_{n}$  which sits inside the positive region of kinematic space.\footnote{Positive region of kinematic space is defined by $X_{ij}\ \geq\ 0, \forall\ i,j$.}
 In fact, our definition of this kinematic Stokes polytope will be such that it is located inside the kinematic associahedron ${\cal A}_{n}$, thus ensuring that it lies in the positive region.  

For $Q_{1}\ =\ (14)$  the $Q_{1}$ compatible set is given by $\{(14, +),\ (36,-)\}$. The corresponding Stokes polytope is one dimensional with two vertices. We locate this Stokes polytope inside the kinematic space via the following constraints. 

\begin{equation}\label{n=6constraint1}
\begin{array}{lll}
s_{ij}\ =\ - c_{ij} \quad  \forall\ 1 \le i  <  j  \le n-1=5 , \ |i-j| \geq 2 \\
X_{13}\ =\ d_{13},\ X_{15}\ =\ d_{15}, \textrm{with}\ d_{13},\ d_{15}\ >\ 0
\end{array}
\end{equation}

The first line of constraints are precisely the ones which define the three dimensional kinematic associahedron ${\cal A}_{6}$ inside ${\cal K}_{6}$. We have motivated the remaining two constraints as follows. We can adjoin, to the diagonal $(14)$ any one out  of the following pairs.\\
${\cal I}\ =\ \{(13,\ 15),\ (24,\ 15),\ (13,\ 46),\ (24,46)\}$ to form a complete triangulation of the hexagon. We pick \emph{any one} of these pairs to impose further constraints on the kinematic data. From the perspective of Feynman diagrams, these constraints are rather natural as planar variables from this set can never occur in Feynman diagrams of $\phi^{4}$ theory. 

Using the above constraints, it can be easily checked that the planar kinematic variables satisfy,

\begin{equation}
\begin{array}{lll}
X_{36}\ =\ -X_{14}\ + c_{14}\ +\ c_{24} + c_{15}\ +\ c_{25}\ \geq 0\\
X_{25}\ =\ d_{15}\ +\ c_{14}\ -\ d_{13}\ +\ c_{13}\ \geq 0
\end{array}
\end{equation}

We thus see that we have a (one dimensional) Stokes polytope ${\cal S}^{Q=(14)}_{6}$ whose vertices are given by $X_{14}\ =\ 0$ and $X_{36}\ =\ 0$ (which is when $X_{14}\ =\ c_{14}\ +\ c_{24}\ +\ c_{15}\ +\ c_{25}$) which correspond to the two $Q$-compatible quadrangulations. It can be readily verified that the kinematic Stokes polytope is insensitive to which of the pairs of diagonals in ${\cal I}$ above we choose to constrain. 
We can now pull back the form given in eqn. (\ref{1formfor6}) on ${\cal S}_{6}$

\begin{equation}\label{spform2}
\begin{array}{lll}
\omega^{Q_{1}}_{6}\ =\ \left(\frac{1}{X_{14}}\ +\ \frac{1}{X_{36}}\right) d X_{14}
=:\ m_{6}({\cal S}^{Q_{1}}_{6})\ d X_{14}
\end{array}
\end{equation}

$m_{6}(Q_{1})$ is the canonical rational function associated to the Stokes polytope ${\cal S}^{Q_{1}}_{6}$. We will use this notation through out the paper namely, we will denote a canonical rational function associated to a Stokes poytope ${\cal S}_{n}^{Q}$ as $m_{n}(Q)$.

As a one dimensional Stokes polytope is also an associahedron (see appendix  \ref{appendixA}), and as the form in eqn.(\ref{spform2}) is the canonical form on associahedron, we have a canonical form on ${\cal S}_{6}^{Q=(14)}$. 

The rational function $m_{6}$ \footnote{For the sake of pedagogy,  we are not differentiating between reference quadrangulation $Q$ that we fix which is in rotated (blue) polygon and quadrangulations which generate stokes polytope which are quadrangulations of the red polygon \cite{baryshnikov:hal-01197226}.} is 

\begin{equation}
m_{6}(Q_{1})\ =\ \left(\frac{1}{X_{14}}\ +\ \frac{1}{X_{36}}\right)
\end{equation}

We can now repeat the analysis with $Q_{2}\ =\ (25)$ and $Q_{3}\ =\ (36)$ analogously and it can be shown that the corresponding canonical forms on the Stokes polytopes are,

\begin{equation}
\begin{array}{lll}
\omega^{Q_{2}}_{6}\ =\ \left(\frac{1}{X_{25}}\ +\ \frac{1}{X_{14}}\right) d X_{25}\\
\omega^{Q_{3}}_{6}\ =\ \left(\frac{1}{X_{36}}\ +\ \frac{1}{X_{25}}\right) d X_{36}
\end{array}
\end{equation}

We now define a function $\widetilde{{\cal M}}_{n}$ on the kinematic space which is a weighted sum of the $m_{6}$ over all ${\cal S}_{n}^{Q}$. In the $n=6$ case this function is defined as,

\begin{equation}\label{n=6sum}
\begin{array}{lll}
\widetilde{{\cal M}}_{6}\ :=\ \alpha_{Q_{1}}\left(\frac{1}{X_{14}}\ +\ \frac{1}{X_{36}}\right)\ +\ \alpha_{Q_{2}}\left(\frac{1}{X_{25}}\ +\ \frac{1}{X_{14}}\right)\ +\ \alpha_{Q_{3}} \left(\frac{1}{X_{36}}\ +\ \frac{1}{X_{25}}\right)
\end{array}
\end{equation}

Here $\alpha_{Q_{i}}$ are positive constants. It is immediately evident that  if and only if $ \alpha_{Q_{1}}\ =\ \alpha_{Q_{2}}\ =\ \alpha_{Q_{3}}\ =\ \frac{1}{2} $,  $ \widetilde{{\cal M}}_{6}\ =\ {\cal M}_{6} $. 
\subsection{Eight particle scattering}

Let us now consider the $n=8$ case. 

Our analysis will proceed along the same lines as in the previous section. Namely we first define planar scattering form on ${\cal K}^{Q}_{8}$ for all the quadrangulations. We will then show how  all the kinematic Stokes polytopes ${\cal S}_{8}^{Q}$  sit inside the 5 dimensional associahedron ${\cal A}_{8}$ and then show how a weighted sum of canonical rational functions over all the polytopes leads to the planar scattering amplitude. 
 
This computation can be made much easier by realising that all the quadrangulations of an octagon (and in general any polygon) can be obtained from cyclic permutations of a subset of quadrangulations. We call this set, set of primitive quadrangulations. More in detail,

Given a $n$ sided polygon with labelled vertices, we call a set of quadrangulations $\{Q_{1},\dots, Q_{I}\}$ primitive if,\\
(a) no two members of the set are related to each other by cylic permutations and\\
(b) all the other quadrangulations can be obtained by a (sequence of) cyclic permutations of one of the $Q$s belonging to the set.

We note that, choice of which quadrangulations are called primitive is not unique but the cardinality of the set of primitive quadrangulations is uniquely fixed by $n$.  In the $n=6$ case, there is only one primitive $Q$ and can be chosen to be $Q\ =\ (14)$. 
 
As shown in section \ref{q-comp}, there are two primitive $Q$'s in this case. 
With out loss of generality we can take them to be $\{Q\ =\ (14,58),\ Q^{\prime}\ =\ (14,16)\}$. 

 As we have shown in figure \ref{Stokes-poly}, \\
  $Q$ compatible quadrangulations are given by : $S_{1}\ =\ \{ (14,58; +),\ (14,47; -)\ (83,58; -)\ (83,47; +)\}$,  \\ 
$Q^{\prime}$ compatible quadrangulations are :  $S_{2}\ =\ \{(14,16; +)\ (14,58;-)\ (36,16;-)\ (36,83;+)\ (58,83;-)\}$.  The signs associated to each quandrangulation is obtained by measuring the number of relative flips from the reference $Q$.\footnote{It is important to maintain the order of the diagonals when a flip is taken as these denote the ordering of the wedge product ($(14, 58) \rightarrow d \ln X_{14} \wedge d \ln X_{58}$ etc.) and since this also contributes to the overall sign of the term when the Scattering form is written down.}

Using eqn. \eqref{psf}, for each of the two sets $S_{1},\ S_{2}$ we can define two distinct planar 2-forms on ${\cal K}_{8}$ as, 
\begin{equation}
\begin{array}{lll}
\Omega^{Q}_{8} =  \left(d\ln X_{14}\ \wedge d\ln X_{58} +\ d\ln X_{38}\ \wedge\ d\ln X_{47}\ -\ d\ln X_{14}\wedge d\ln X_{47}\ - d\ln X_{38}\ \wedge d\ln X_{58}\ \right) \\
\Omega^{Q^{\prime}}_{8}\ =\ 
\left(\ d\ln X_{14}\wedge\ d\ln X_{16}\ -\ d\ln X_{14}\wedge d\ln X_{58}\ -\ d\ln X_{36}\wedge d\ln X_{16}\ \right.\\ \hspace*{3.67in} \left. + \ d\ln X_{36}\wedge d\ln X_{83}\ - d\ln X_{58}\wedge d\ln X_{83}\ \right)
\end{array}
\end{equation}

One can write down scattering forms for all other quadrangulations exactly analogously. The Stokes polytopes associated to $S_{1}$, $S_{2}$ are two dimensional positive geometries with four and five vertices respectively. 

We now locate the two Stokes polytopes $S_{Q}$ and $S_{Q^{\prime}}$ inside the Kinematic space (in fact, inside the five dimensional associahedron ${\cal A}_{8}$) precisely in analogy with $n = 6$ case. 
Let $T_{1}$ and $T_{2}$ be \emph{any} two sets of diagonals which are such that $T_{1} \cup  \{14,58\}$ and $T_{2} \cup \{14,16\}$ are complete triangulations of the octagon (with labelled vertices). We choose $T_{1}$ and $T_{2}$ to be $\{13,48,57\}$ and $\{13,46,86\}$ respectively.\footnote{As can be easily verified by the reader, any of the other 8 allowed choices of $T_{1},\ T_{2}$ will also suffice.}

The constraints defining $S_{Q_{1}}$ and $S_{Q_{2}}$ inside the kinematic space are respectively given by 

\begin{equation}
\begin{array}{lll}
s_{ij}\ =\ -c_{ij}\ \forall\ 1\leq\ i\ <\ j\ \leq\ 7\ \textrm{with}\ \vert i - j\vert\ \geq 2\\
X_{13}\ =\ d_{13},\ X_{48}\ =\ d_{48}\ , X_{57}\ =\ d_{57}
\end{array}
\end{equation}

\begin{equation}
\begin{array}{lll}
s_{ij}\ =\ -c_{ij}\ \forall\ 1\leq\ i\ <\ j\ \leq\ 7\ \textrm{with}\ \vert i - j\vert\ \geq 2\\
X_{13}\ =\ d_{13},\ X_{46}\ =\ d_{46}\ , X_{68}\ =\ d_{68}
\end{array}
\end{equation}

These constraints locate both the Stokes polytopes inside the five dimensional associahedron ${\cal A}_{8}$ and hence ensure that all the $X_{ij}$'s are positive in the interior of the Stokes polytopes.

Using these constraints it is simple algebraic exercise to show that on ${\cal S}^{Q}_{8}$, ${\cal S}^{Q^{\prime}}_{8}$ one has the following top forms obtained from $\Omega_{Q_{i}}$ on ${\cal K}_{8}$. 

\begin{equation}
\begin{array}{lll}
\omega_{8}^{Q}\ =
\left(\frac{1}{X_{14}X_{58}}\ +\ \frac{1}{X_{38}X_{47}}\ +\ \frac{1}{X_{14}X_{47}}\ +\ \frac{1}{X_{38}X_{58}}\
\right)\ d X_{14}\ \wedge d X_{58}\\
\omega_{8}^{Q^{\prime}}\ =
\left(\ \frac{1}{X_{14}X_{16}}\ +\ \frac{1}{X_{14}X_{58}}\ +\ \frac{1}{X_{36}X_{16}}\ +\ \frac{1}{X_{36}X_{83}}\ +\ \frac{1}{X_{58}X_{83}}\ \right)\ dX_{14}\ \wedge\ d X_{16}
\end{array}
\end{equation}


The corresponding canonical functions $m_{8}$ are given by 

\begin{equation}
\begin{array}{lll}
m_{8}(Q)\ =\ \left(\frac{1}{X_{14}X_{58}}\ +\ \frac{1}{X_{38}X_{47}}\ +\ \frac{1}{X_{14}X_{47}}\ +\ \frac{1}{X_{38}X_{58}}\
\right)\\
m_{8}(Q^{\prime})\ =\ \left(\ \frac{1}{X_{14}X_{16}}\ +\ \frac{1}{X_{14}X_{58}}\ +\ \frac{1}{X_{36}X_{16}}\ +\ \frac{1}{X_{36}X_{83}}\ +\ \frac{1}{X_{58}X_{83}}\ \right)
\end{array}
\end{equation}

As all the other quadrangulations can be obtained by cyclic permutations of (labels of) $Q$ and $Q^{\prime}$, we can easily write down the functions $f$ associated to all the Stokes polytopes and substitute them in $\widetilde{{\cal M}}_{8}$ 

\begin{equation}\label{master`}
\widetilde{{\cal M}}_{8}\ =\ \sum_{\sigma}\alpha_{\sigma\cdot Q}\ m_{8}({\sigma\cdot Q})\ +\ \Sigma_{\sigma^{\prime}}\alpha_{\sigma^{\prime}\cdot Q^{\prime}}\ m_{8}({\sigma^{\prime}\cdot Q^{\prime}})
\end{equation}

where $\sigma, \sigma^{\prime}$ range over all the cyclic permutations which map $Q$ and $Q^{\prime}$ to distinct quadrangulations respectively. 

Upon substituting the residues in eqn. \eqref{master`}, it can be easily checked that there is a unique choice of $\alpha$ s , namely $\alpha_{\sigma\cdot Q}\ =\ \frac{2}{6}\ \forall\ \sigma$ and $\alpha_{\sigma^{\prime}\cdot Q^{\prime}}\ =\ \frac{1}{6}\ \forall \sigma^{\prime}$, for which $\widetilde{{\cal M}}_{8}\ =\ {\cal M}_{8}$ (see appendix \ref{appendixB}).

\section{Computing ${\cal M}_{n}$ from the canonical forms}\label{primitives}

As we saw in the previous section, in both the $n=6$ and $n=8$ cases the scattering amplitude can be obtained from a weighted sum of rational functions (associated to canonical forms) over all the Stokes polytopes. A curious fact about the weights $\alpha$ was that the  $\alpha$ s for which $\widetilde{{\cal M}}_{n}$ equals ${\cal M}_{n}$ were parametrized only by the primitive quadrangulations. In other words, in both the cases considered above, 
\begin{equation}
\alpha_{Q}\ =\ \alpha_{\sigma\cdot Q}\ \forall\ \sigma
\end{equation}

We also formalize this observation into a constraint on the weights as
\begin{equation}\label{constraint}
\alpha_{Q}\ =\ \alpha_{Q^{\prime}}\ \textrm{if $Q^{\prime}\ =\ \sigma\cdot\ Q$ for a cyclic permutation $\sigma$}
\end{equation}

That is if two quadrangulations are related by a cylic permutation of vertices of the polygon, then the corresponding $\alpha$ s should be equal. 

The underlying motivation for the constraint in \eqref{constraint} is the following. Consider two quadrangulations $Q$ and $Q^{\prime}$ which are cyclically related.  From the perspective of kinematic Stokes polytope this means that the difference between ${\cal S}_{Q^{\prime}}$ and ${\cal S}_{Q}$ is simply in how they are embedded in the kinematic space. Our constraints are based on our intuition (based on $n=6,\ 8$ cases) that $\alpha_{Q}$ only depend on the intrinsic (combinatorial) property of ${\cal S}^{Q}$ and not on how it is embedded in ${\cal K}_{n}$. This dependence of $\alpha$'s on certain equivalence class of quadrangulations can be encapsulated by the notion of primitive quadrangulations.

We now propose a formula
for evaluating the function $\widetilde{{\cal M}}_{n}$ for arbitrary $n$.   

\begin{equation}\label{master}
\begin{array}{lll}
\widetilde{{\cal M}}_{n}\ =\displaystyle \ \sum_{Q\vert\textrm{primitive}}\ \sum_{\sigma}\ \alpha_{Q}\ m_{n}(\sigma\cdot Q)
\end{array}
\end{equation}

The proposal (for computing the planar scattering amplitude ${\cal M}_{n}$) can thus be summarised as follows :  For any $n$ we first compute $m_{n}({\sigma\cdot Q})$ and substitute in eqn. \eqref{master}. We conjecture that there is a unique choice of $\alpha$s 
\emph{which should be computed purely from combinatorics of $Q$ s} such that for these $\alpha$ s, 
$\widetilde{{\cal M}}_{n}\ =\ {\cal M}_{n}$. That is, there is a unique choice of $\alpha_{Q}\ \forall\ \textrm{primitive}\ Q$ such that contribution of all the poles to $\widetilde{{\cal M}}_{n}$ with residue unity.

We should emphasize that to compute the scattering amplitude ${\cal M}_{n}$ from residues of the Stokes polytopes, we need an independent formula for $\alpha_{Q}$ which is consistent with eqn. \eqref{constraint}, and such that all the kinematic channels give equal contribution of order unity.  We do not have such a formula so far and in this paper, we have attempted to verify this formula in a handful of examples. In appendix \ref{appendixB} we verify that our proposal leads to the correct scattering amplitude for ten point scattering amplitude.

We also emphasise that our formula is a mere repackaging of the ``more fundamental'' formula

\begin{equation}
\widetilde{{\cal M}}_{n}\ =\ \sum_{Q}\alpha_{Q}\ m_{n}(Q)
\end{equation}

where one sums over all the Stokes polytopes (parametrized by $Q$), with the proviso that $\alpha_{Q}$ are same for any two quadrangulations which are related by cyclic permutation. 

It is important to summarise our story so far. We have shown that given any quadrangulation $Q$ of an $n$-sided polygon, one can define a unique planar scattering form on the kinematic space ${\cal K}_{n}$. We then showed how this form naturally descends to the canonical form on the Stokes polytope ${\cal S}_{n}^{Q}$ such that the corresponding rational function $m_{n}$ gives a partial contribution to planar scattering amplitude in $\phi^{4}$ theory. Thus an individual Stokes polytope is not quite the same as an amplituhedron which as a single geometric object contained information about complete scattering amplitude. 
However the families of all Stokes polytope does contain  complete information about ${\cal M}_{n}$. We proposed a formula for obtaining ${\cal M}_{n}$ by summing over $m_{n}(Q)$ of all the Stokes polytopes and have shown it to be valid for 6, 8 and 10 particle amplitudes. It is important to stress that a single Stokes polytope is not the amplituhedron of planar amplitudes in massless $\phi^{4}$ theory.

\section{Factorization}\label{factorization}

One of the remarkable consequences of relating tree level scattering amplitudes to  positive geometries like associahedron is the fact that geometric factorization of the associahedron implied physical factorization of scattering amplitude. This in turn implied that tree-level unitarity and locality are emergent properties of the positive geometry \cite{Arkani-Hamed:2017mur}. In this section we will try to argue that this is indeed the case even for planar amplitudes in  massless $\phi^{4}$ theory. Namely that, there is a combinatorial factorization of Stokes polytope and that exactly as in the case of associahedron, it implies amplitude factorization. 

Our first assertion is the following. Given any diagonal $(ij)$,  consider \emph{all} $Q$ which contains $ij$ and the consider all the corresponding kinematic Stokes polytopes ${\cal S}_{n}^{Q}$. We contend that for each of these Stokes polytopes, the corresponding facet $X_{ij}\ =\ 0$ is a product of lower dimensional Stokes polytopes.  
\begin{equation}\label{factor1}
{\cal S}_{n}^{Q}\bigg|_{X_{ij}\ =\ 0}\ \equiv\ {\cal S}_{m}^{Q_{1}}\ \times\ {\cal S}_{n+2-m}^{Q_{2}}
\end{equation}

where $Q_{1}$ and $Q_{2}$ are such that $Q_{1}\ \cup Q_{2}\ \cup\ (ij)\ =\ Q$. $Q_{1}$ is the quadrangulation of the polygon $\{ i,\ i+1,\dots,\ j\}$  and $Q_{2}$ is the quadrangulation of $\{j,j+1,\dots,n,\dots,i\}$. 
Now we know that, on ${\cal S}_{n}^{Q}$ any planar scattering variable $X_{kl}$ is a linear combination of $X_{ij}$ and remaining $X$'s which constitute $Q$. Hence in order to prove this assertion we need to show that  any $X_{kl}$ with $ i \leq k  <  l  \leq j$ can be written as a linear combination of $X_{ij}$ and elements of $Q_{1}$ and  similarly any variable in the complimentary set can be written in terms of $X_{ij}$ and elements of $Q_{2}$.

However this is immediate since we know from the factorization property of associahedron proven in \cite{Arkani-Hamed:2017mur} that any $X_{kl} = \displaystyle \ X_{ij}\ + \sum_{i < m < n < j} X_{mn}$. some of these $X_{mn}\ \in\ Q_{1}$ and the others are constrained via $X_{mn} = d_{mn}$. This proves our assertion. Thus $X_{ij} = 0$ facet factorizes into two lower dimensional Stokes polytopes. 

Our second assertion is that the geometric factorization implies amplitude factorization of quartic theory. This assertion is based on the following two facts.\\
(1) As Stokes polytope is a positive geometry , we know that it's canonical form satisfies the following properties satisfed by canonical form on any positive geometry ${\cal A}$ (For details, we refer the reader to appendix A of \cite{Arkani-Hamed:2017mur} and \cite{Arkani-Hamed:2017tmz}). 

\begin{equation}
\textrm{Res}_{H} \, \omega_{{\cal A}}\ =\ \omega_{{\cal B}}
\end{equation}

where we think of $\omega_{{\cal A}}$ as defined on the embedding space and $H$ is any subspace in the embedding space which contains the face ${\cal B}$.
It is also known that if ${\cal B}\ =\ {\cal B}_{1}\times{\cal B}_{2}$ then

\begin{equation}
\omega({\cal B})\ =\ \omega({\cal B}_{1})\, \wedge\, \omega({\cal B}_{2})
\end{equation}

Thus we immediately see that 

\begin{equation}
\textrm{Res}_{\, X_{ij}\, =\, 0} \;\; \omega({\cal S}_{n}^{Q})\ =\ \omega_{m}^{Q_{1}}\, \wedge\, \omega_{n+2-m}^{Q_{2}}\ \forall\ Q. 
\end{equation}
where $m\ =\ j - i + 1$. 

We thus see that residue over each Stokes polytope which contains a boundary $X_{ij}\ \rightarrow\ 0$ factorizes into residues over lower dimensional Stokes polytopes. 
This factorization property naturally implies factorization of amplitudes as follows. Consider the $n$-gon with a diagonal $(ij)$ (with $i,j$ such that this diagonal can be part of a quadrangulation). This diagonal subdivides the $n$-gon into a two polygons with vertices $\{i,\dots,j\}$ and $\{j,\dots,n,1,\dots i\}$ respectively.  By considering all the kinematic Stokes polytopes associated to these polygons, we can evaluate  $\widetilde{M}_{\vert j - i + 1\vert},\ \widetilde{M}_{n+2 - (\vert j - i + 1\vert)}$ which correspond to left and right sub-amplitudes respectively. This immediately implies that 

\begin{equation} \label{factor2}
\widetilde{{\cal M}}_{n}\bigg|_{X_{ij}\, =\, 0}\ =\ \widetilde{{\cal M}}_{L} \,\frac{1}{X_{ij}}\, \widetilde{{\cal M}}_{R}
\end{equation}

This proves physical factorization. We also note that, eqns. (\ref{factor1}) and (\ref{factor2}) imply following constraints on $\alpha$ s.

\begin{equation} \label{factor3}
\sum_{Q\ \textrm{containing} (ij)} \alpha_{Q}\ =\ \sum_{Q_{L}, Q_{R}}\alpha_{Q_{L}}\alpha_{Q_{R}}
\end{equation}

where $Q_{L}$ and $Q_{R}$ range over all the quadrangulations of the two polygons to the left and right of diagonal $(ij)$ respectively.

It can be verified that in the case of $n\ =\ 6,8,\ \textrm{and}\ 10$ particles $\alpha_{Q}$'s do indeed satisfy these constraints\footnote{We expect eqn. (\ref{factor3}) to be useful in determining $\alpha$ s.}.

 \section{Relationship with planar scattering form for cubic coupling}\label{cubicquartic}
 
Planar tree-level diagrams of  massless $\phi^{4}$ theory can be obtained from diagrams of a theory with cubic interactions $\psi\phi^{2}$ which contains two scalar fields $\phi$ and $\psi$, where $\phi$ is massless and $\psi$ is massive. Consider an (ordered) $n$-point amplitude in this theory ${\cal M}^{\phi^{2}\psi}(p_{1},\dots, p_{n})$ in which all the external particles are $\phi$-particles. The super-script on the amplitudes indicates the coupling we are considering. It is easy to see that in all the Feynman graphs associated to such an amplitude, the $\phi$-propagators  precisely correspond to the $\phi$-propagators in the corresponding diagrams in $\phi^{4}$ theory. Remaining propagators are propagators associated to $\psi$ field and hence upon  integrating out this  massive field,  one recovers planar amplitudes in massless $\phi^{4}$ theory. 
 
Whence one may wonder if the canonical form we obtained on  Stokes polytopes, ${\cal S}^{Q}_{n}$ could be obtained  from the planar scattering form associated to the theory with $\psi\phi^{2}$  interaction. \footnote{We are indebted to Nemani Suryanarayana and Suresh Govindarajan for raising this question. We also note that this issue was already raised in \cite{Arkani-Hamed:2017mur}.} We show below that this is not the case. 

We can postulate a planar scattering form in the kinematic space associated to $\psi\phi^{2}$ coupling, in which  all the log singularities associated to $\psi$ fields are absent\footnote{This is how we implement ``integrating out the $\psi$-field" in language of scattering forms.}. On restricting this form to ${\cal S}^{Q}_{n}$, we can observe that the corresponding form is not the canonical form on ${\cal S}^{Q}_{n}$. 

Let us illustrate this idea in the simplest of examples, namely $n = 6$ case.  We thus consider planar scattering form on ${\cal K}_{6}$ which is obtained by summing over 12 planar graphs\footnote{In the case of $\phi^{3}$ coupling, one has to sum over 14 graphs, however two of these do not arise if we instead consider $\psi\phi^{2}$ coupling. Whence the corresponding form on ${\cal K}_{6}$ is not projective! In the context of triangulation, what this means is that we consider only those triangulations which has \emph{at least} one partial triangulation which can be part of a quadrangulation.}.

 This form is given by

\begin{equation}
\begin{array}{lll}
\Omega^{\psi\phi^{2}}_{n=6}\ =\\
d X_{24}\ \wedge\ d\ln X_{14}\ \wedge\ [d X_{15}\ -\ d X_{46} ]\ +\ d X_{26}\wedge d\ln X_{36}\ \wedge\ [d X_{46}- d X_{35}]\\
-\ d X_{13}\wedge d\ln X_{36}\ \wedge [\ d X_{46} - d X_{35}\ ]\ -\ d X_{26}\wedge d\ln X_{25}\wedge [\ d X_{24}\ -\ d X_{35}\ ]\\
+\ d X_{15}\wedge d\ln X_{25}\ \wedge [\ d X_{24}\ -\ d X_{35}\ ]\ -\ d X_{13}\ \wedge d\ln X_{14}\ \wedge [d X_{15} - d X_{46}\ ]
\end{array}
\end{equation}

where singularities associated to $\psi$ propagators are absent.

On restricting this form to ${\cal S}_{6}^{Q = (14)}$ using eqn. (\ref{n=6constraint1}), we get

\begin{equation}\label{yukawa}
\tilde{\Omega}_{N=6}\bigg|_{{\cal S}^{Q=(14)}_{6}}\ =\ 2\ \left[\frac{1}{X_{14}}\ +\ \frac{1}{X_{25}}\ +\ \frac{1}{X_{36}}\right]\ d X_{13}\ \wedge d X_{14}\ \wedge d X_{15}
\end{equation}

We thus see that projection of $\Omega^{\psi\phi^{2}}_{n=6}$ onto ${\cal S}^{Q=(14)}_{6}$ is not the same as its canonical form. This is because the form in eqn.\eqref{yukawa} has an additional singularity at $X_{25}\ \rightarrow\ 0$. Thus from the perspective of positive geometry there does not seem to be a direct relationship between quartic interactions and cubic interactions with two scalar fields. Of course in hindsight, this is not too surprising as integrating out the $\psi$ field reproduces all (planar) diagrams in $\phi^{4}$ theory and this is precisely reflected in the presence of $\frac{1}{X_{25}}$ in eqn. \eqref{yukawa} above. However as the $X_{25}\ \rightarrow\ 0$ singularity is not on one of the vertices of the Stokes polytope, this form is not the canonical form on the Stokes polytope.  We leave further investigation of relationship between cubic and quartic couplings in the context of positive geometries for future work. 

\section{Conclusion}

The connection between differential forms in kinematic space, polytopes and scattering amplitudes is unravelling a deeper structure of quantum field theories by unifying several recent developments like color-kinematics duality, Recursion relations and CHY formula into one theme \cite{Arkani-Hamed:2017mur, Gao:2017dek, He:2018svj}. For tree level scattering amplitudes in a variety of theories, these multi-faceted connections are precise and centre around combinatorial geometry of the polytope in the kinematic space. For planar diagrams in scalar field theory with cubic coupling, this polytope is  a well known classic polytope , associahedron.  In this paper, we have tried to explore these connections in the context of massless $\phi^{4}$ theory and shown that the connections continue to hold, although with several caveats. 

As we saw above, there is no single polytope which encompasses all the information about the scattering amplitude. There is a family of polytopes each of whose combinatorial geometry contains partial information about the amplitude in such a way that a weighted sum over all the Stokes polytopes produces complete scattering amplitude. Our analysis is rather nascent but opens several interesting avenues for further investigations. 

There is first an obvious unsolved issue of computing the weights.  In order to give a formula where scattering amplitude is completely determined by combinatorial geometry of Stokes Polytopes, a formula for the weights $\alpha_{Q}$ should be derived. Our contention, based on several examples is that these weights only depend on combinatorics of the so-called primitive quadrangulations. However a formula for the weights is missing so far. 

There is also an obvious question of how to go beyond planar amplitudes and is there a polytope realisation for full tree-level scattering amplitude of $\phi^{4}$ theory. In the massless $\phi^{3}$ case, certain progress in this direction was already reported in \cite{Arkani-Hamed:2017mur, Gao:2017dek}. It was shown that a wider class of amplitudes then simply planar ones could be computed with the corresponding polytopes being generalisation of associahedra known as Cayley polytopes. It will be interesting to see if by generalising Stokes polytopes (to more general polytopes associated with quadrangulations) we can go beyond planar diagrams in $\phi^{4}$ theory.

One of our central motivations for this work was to see if the CHY integrand for (planar diagrams) in $\phi^{4}$ theory can also be understood as pull-backs of certain forms on kinematic space. It is here that a fascinating question emerges. In the world of CHY formalism, $n$-particle tree-level scattering amplitude for \emph{any} theory containing massless particles  is a result of integrating a top-form on worldsheet moduli space. Hence this form is always $(n-3)$-form. But as the dimension of Stokes polytope is $\frac{n-4}{2}$, we see that pullback of such a form (using scattering equations) onto the worldsheet moduli space will not be a top form. Such lower forms have not played a role in CHY formalism so far and it will be interesting to unravel this connection clearly. Going in the other direction, if we push forward the CHY top-form for planar $\phi^{4}$ theory onto kinematic space, one would get a $(n-3)$ form on ${\cal K}_{n}$ and it will be interesting to explore the relationship of this form with the canonical form on Stokes polytope.\footnote{We are indebted to Song He for discussions on this point.}

We believe that our work can be generalised to planar diagrams in $\phi^{p},\ p\ >\ 4$ interactions. The notion of Q-compatible quadrangulations which formed the vertices of the Stokes polytope has an immediate extension to $p$-gulations of a polygon. In the case of lower point amplitudes (say $n=10$ particle scattering in $\phi^{6}$ case) , it can be checked that our analysis admits a step-by-step generalisation and produces weighted sum  over certain (hitherto unknown) polytopes which for certain choice of weights yield the scattering amplitude. 

\section*{Acknowledgement}

We are extremely thankful to Sujay Ashok for numerous discussions on various issues related to this work, for his constant encouragement and especially for discussions on world-sheet associahedron. PB is thankful to ICTS string theory group particularly to the participants of Student/Postdoc Journal Club for stimulating discussions on related topics. AL is indebted to Song He for his patient explanations related to the amplituhedron program  and raising several important points which significantly improved our understanding. PR would like to thank Nemani Suryanarayana for many discussions and several prescient remarks over the course of this work. We would like to thank participants of Chennai Strings Meeting, especially Suresh Govindarajan for their inputs.  AL is grateful to Institute of Theoretical Physics, Beijing for their hospitality where part of this work was done. This research was supported in part by the International Centre for Theoretical Sciences (ICTS) during a visit for participating in the program - Kavli Asian Winter School (KAWS) on Strings, Particles and Cosmology 2018 (Code: ICTS/Prog-KAWS2018/01). AL and PR would like to thank ICTS, where this work began, for the hospitality.

\appendix

\section{\label{appendixA}Few facts about Stokes polytopes} 

\begin{figure}[H]
\centering
\includegraphics[width=10cm]{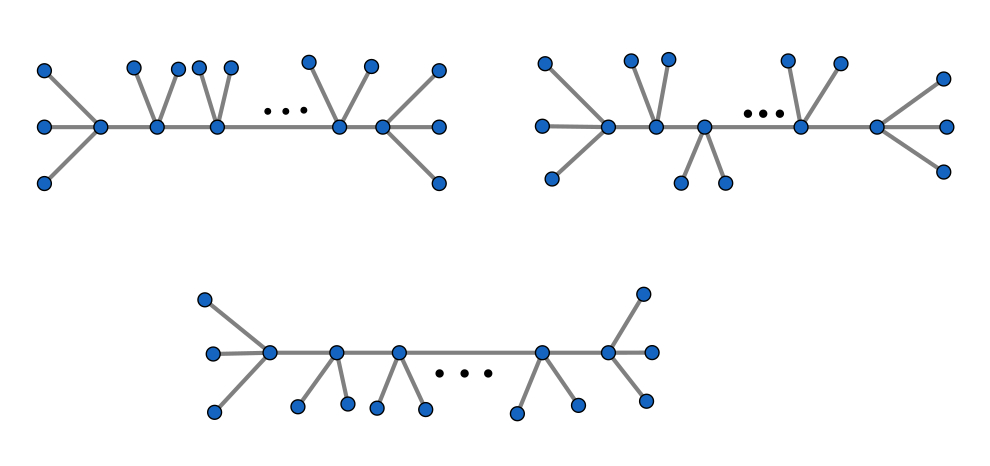}
\caption{All the diagrams that have associahedra as their polytope.}\label{snakechannels}
\end{figure}

In this appendix we will review some known facts about Stokes polytopes that may help in understanding the maintext better. We will be mainly following \cite{ Baryshnikov, chapoton, baryshnikov:hal-01197226}.
\begin{itemize}
\item Whenever we encounter a $\large{+}$ junction as in figure \ref{fig: splitting} then the corresponding Stokes polytope splits into product of lower dimensional Stokes polytopes\footnote{We would like to emphasize that for the diagrams themselves there is no such splitting only the corresponding Stokes polytope splits.}.
\begin{figure}[H]
\centering
\includegraphics[width=12cm]{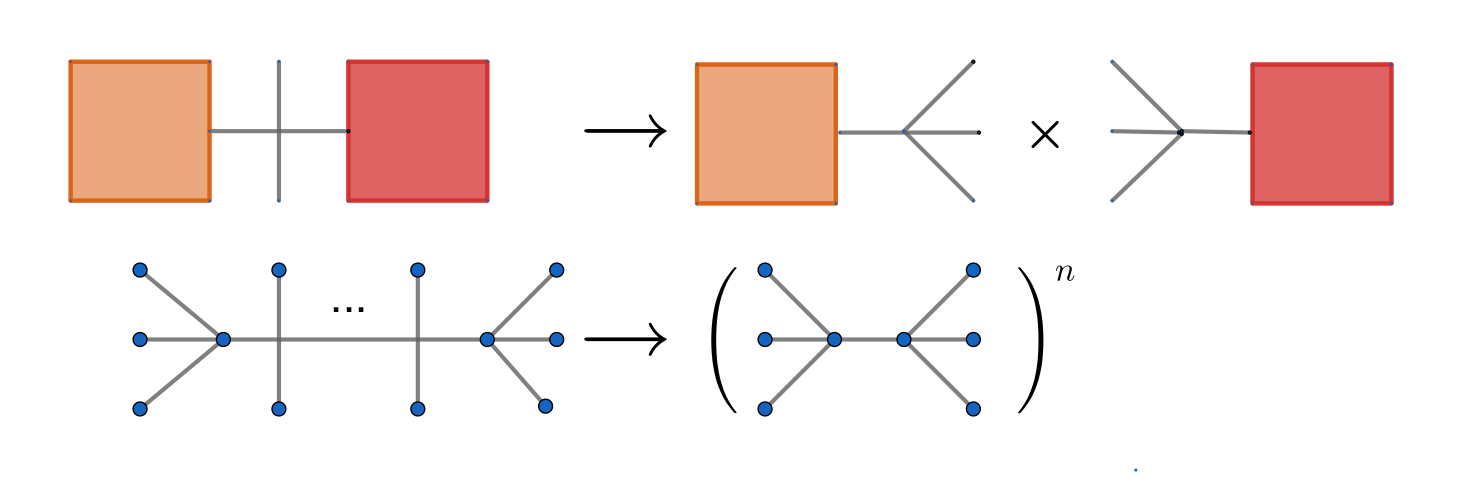} 
\caption{The upper figure shows the splitting into lower dimensional polytopes. The lower figure shows explains why the Stokes polytope for this case is a hyper-cube.}\label{fig: splitting}
\end{figure}

\noindent This is easy to see as if the reference quadrangulation is given by $Q= \{i_1 j_1, i_2 j_2,\cdots, i_n j_n \}$ with $\{i_1 j_1, i_2 j_2,\cdots, i_{k-1} j_{k-1} \}$ and $\{i_{k+1} j_{k+1},\cdots, i_n j_n \}$ denoting the left and right half of the quadrangulation respectively, we could perform flips in each half independently to get all the vertices of the Stokes polytope as regardless of the flip the diagonals of the two halves  never enter $\{i_{k-1}, i_k ,i_{k+1}, j_{k-1}, j_k ,j_{k+1}\}$.

\item When we twist a quartic graph about any propagator as in figure \ref{fig: Twisting}, the corresponding Stokes polytope does not change.

\begin{figure}[H]
\centering
\includegraphics[width=12cm]{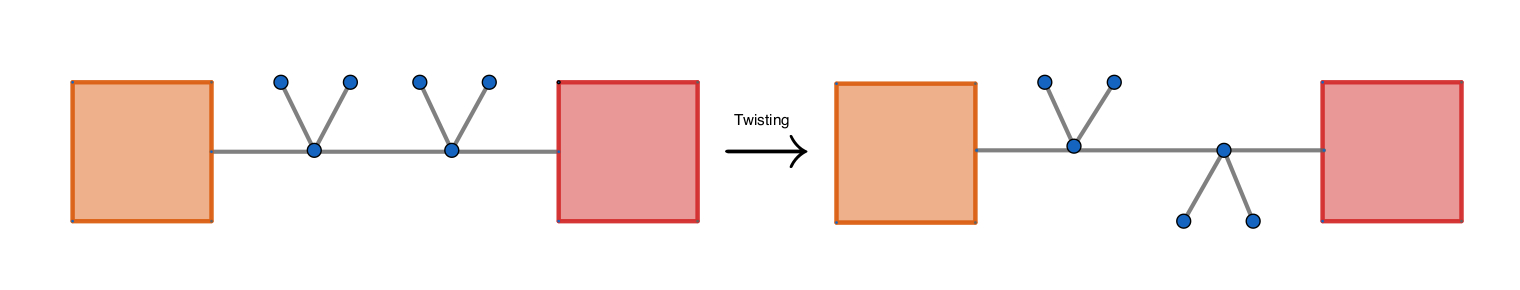} 
\caption{A quartic graph before and after twisting.}\label{fig: Twisting}
\end{figure}
This is not so easy to see and needs an introduction of the concept of  certain paths known as serpent nests. We will not attempt to do this here and refer interested reader to \cite{chapoton} for details. This fact does however helps us in understanding why despite there being several topologically inequivalent cubic graphs (corresponding to whether at each vertex the external leg is above or below the central line similar to \ref{snakechannels}) for a given $n$ they all had the same polytope namely the associahedron.

An interesting aspect of a Stokes polytope is the following theorem. 

\item {\bf Theorem}: Any Stokes polytope is writable as a Minkowski sum of hypercubes and associahedra.

By Minkowski sum $M$ of $A$ and $B$ we simply mean\footnote{We can also understand this by treating each point in $A$ and $B$ as the endpoints of a hypothetical vectors so that the resultant belongs to $M$.}:
\begin{equation}
M= \{a+b |a \in A, b \in B\}
\end{equation}
Reader interested in proof of this statement should consult \cite{Baryshnikov}.
Thus, the Stokes polytopes are interpolating polytopes in some sense between the simplest polytope, the cube and the most complicated polytope the associahedron. 

\end{itemize}

There is no known formula for the number of Q-compatible quadrangulations\footnote{This is mainly due to the fact that the Stokes polytopes have not been studied much since their discovery in a different context \cite{Baryshnikov}.} for a generic reference quadrangulation $Q$. However, for a few special quadrangulations such a formula is known and we shall list them below along with the corresponding polytopes. \\

\begin{figure}[H]
	\centering
	\includegraphics[width=6cm]{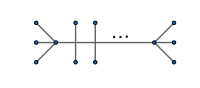} 
	\caption{\label{fig: bridgehypercube}The quartic graph whose corresponding Stokes polytope is the hypercube. The quadrangulation corresponding to this case has only parallel diagonals}
\end{figure}

{\bf 1. Bridge :} This case corresponds to choosing the reference $Q$ for the graph given below in the figure \ref{fig: bridgehypercube}. As explained above the polytope in this case turns out to be a hypercube with $2^{N-1}$ vertices.

\vspace{0.5 cm}

{\bf 2. Snake :} In this case the corresponding polytope is an associahedron with Catalan number $C_{2n-1}$ vertices. There are $2^{N-1}$ such diagrams where $N-1$ is number of vertices. It is easy to see why all of them have the same polytope as they are all related to each other by twisting.

\vspace{0.5 cm}

 {\bf 3. Lucas :} In this case the corresponding polytope has $L_{N-1}$ number of vertices, where \emph{Lucas number} ($L_n$) is defined by the recursion formula : $ L_{0}= 0, \quad L_{1}=2, \quad L_{n+2}= 6 L_{n+1}+ 3 L_{n}.$
\begin{figure}[H]
\centering
\includegraphics[width=6cm]{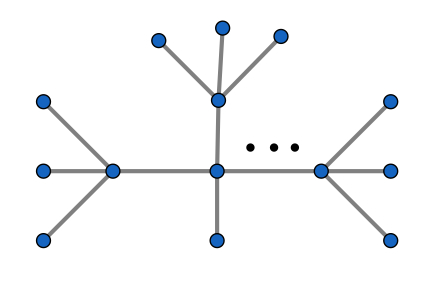}
\caption{The quadrangulation in this case is given by a chain of diagonals $\{i_1j_1,j_1i_2,i_2j_3,...\}$}
\end{figure}
\vspace{0.5cm}
This is not very straightforward to see and the interested reader may find the proof in \cite{baryshnikov:hal-01197226}. In \cite{baryshnikov:hal-01197226} a much more general formula was obtained using the method of generating functions for the case where instead of $\{3,3,n,1\}$ external particles you have $\{n_1,n_2,n_3,n_4\}$ external particles.

\section{\label{appendixB}Some details : For $n=8, 10$ }
\subsection*{Some details of the $n=8$ case}\label{8alpha}
We provide the details of the computation of the $\alpha$ factors for $n=8$ case here. The functions $m_{8}$ corresponding to all $F_4=12$ quadrangulations are given below. There are $4$ Stokes polytopes with 4 vertices and $8$ Stokes polytopes with 5 vertices.
\begin{equation}
\begin{array}{lll}
m_{8}({Q_1})\ =\ \left(\frac{1}{X_{14}X_{58}}\ +\ \frac{1}{X_{38}X_{47}}\ +\ \frac{1}{X_{14}X_{47}}\ +\ \frac{1}{X_{38}X_{58}}\
\right)\\
m_{8}({Q_2})\ =\ \left(\frac{1}{X_{25}X_{16}}\ +\ \frac{1}{X_{25}X_{58}}\ +\ \frac{1}{X_{14}X_{58}}\ +\ \frac{1}{X_{14}X_{16}}\
\right)\\
m_{8}({Q_3})\ =\ \left(\frac{1}{X_{36}X_{27}}\ +\ \frac{1}{X_{36}X_{16}}\ +\ \frac{1}{X_{25}X_{16}}\ +\ \frac{1}{X_{25}X_{27}}\
\right)\\
m_{8}({Q_4})\ =\ \left(\frac{1}{X_{47}X_{38}}\ +\ \frac{1}{X_{47}X_{27}}\ +\ \frac{1}{X_{36}X_{27}}\ +\ \frac{1}{X_{36}X_{38}}\
\right) \nonumber
\end{array}
\end{equation}
\begin{equation}
\begin{array}{lll}
m_{8}({Q^{\prime}_1})\ =\ \left(\ \frac{1}{X_{14}X_{16}}\ +\ \frac{1}{X_{14}X_{58}}\ +\ \frac{1}{X_{36}X_{16}}\ +\ \frac{1}{X_{36}X_{83}}\ +\ \frac{1}{X_{58}X_{38}}\ \right) \\
m_{8}({Q^{\prime}_2})\ =\ \left(\ \frac{1}{X_{25}X_{27}}\ +\ \frac{1}{X_{25}X_{16}}\ +\ \frac{1}{X_{14}X_{16}}\ +\ \frac{1}{X_{47}X_{14}}\ +\ \frac{1}{X_{47}X_{27}}\ \right) \\
m_{8}({Q^{\prime}_3})\ =\ \left(\ \frac{1}{X_{36}X_{38}}\ +\ \frac{1}{X_{36}X_{27}}\ +\ \frac{1}{X_{25}X_{27}}\ +\ \frac{1}{X_{58}X_{25}}\ +\ \frac{1}{X_{58}X_{38}}\ \right) \\
m_{8}({Q^{\prime}_4})\ =\ \left(\ \frac{1}{X_{47}X_{14}}\ +\ \frac{1}{X_{47}X_{38}}\ +\ \frac{1}{X_{36}X_{38}}\ +\ \frac{1}{X_{16}X_{36}}\ +\ \frac{1}{X_{16}X_{14}}\ \right) \\
m_{8}({Q^{\prime}_5})\ =\ \left(\ \frac{1}{X_{58}X_{25}}\ +\ \frac{1}{X_{14}X_{58}}\ +\ \frac{1}{X_{14}X_{47}}\ +\ \frac{1}{X_{27}X_{47}}\ +\ \frac{1}{X_{25}X_{27}}\ \right) \\
m_{8}({Q^{\prime}_6})\ =\ \left(\ \frac{1}{X_{16}X_{36}}\ +\ \frac{1}{X_{16}X_{25}}\ +\ \frac{1}{X_{25}X_{58}}\ +\ \frac{1}{X_{38}X_{58}}\ +\ \frac{1}{X_{36}X_{38}}\ \right) \\
m_{8}({Q^{\prime}_7})\ =\ \left(\ \frac{1}{X_{27}X_{47}}\ +\ \frac{1}{X_{27}X_{36}}\ +\ \frac{1}{X_{16}X_{36}}\ +\ \frac{1}{X_{14}X_{16}}\ +\ \frac{1}{X_{14}X_{47}}\ \right) \\
m_{8}({Q^{\prime}_8})\ =\ \left(\ \frac{1}{X_{38}X_{58}}\ +\ \frac{1}{X_{38}X_{47}}\ +\ \frac{1}{X_{27}X_{47}}\ +\ \frac{1}{X_{25}X_{27}}\ +\ \frac{1}{X_{25}X_{58}}\ \right) \\ \nonumber
\end{array}
\end{equation}
Every term in the above sum has either  $X_{i i+3}X_{j j+3}$ or $X_{i i+3}X_{i i+5}$ in its denominator. We can see that each $X_{i i+3}X_{j j+3}$ term appears twice in the first list and twice in the second list. Similarly, each $X_{i i+3}X_{i i+5}$ term appears only once in the first list and four times in the second list. Thus, we have 
\bea
2 \alpha_{\sigma.Q} + 2 \alpha_{\sigma'.Q'} =1  \nonumber \\
 \alpha_{\sigma.Q} + 4 \alpha_{\sigma'.Q'} =1 \nonumber
\eea
which gives 
$\alpha_{\sigma\cdot Q}\ =\ \frac{2}{6}\ \forall\ \sigma$ and $\alpha_{\sigma^{\prime}\cdot Q^{\prime}}\ =\ \frac{1}{6}\ \forall \sigma^{\prime}.$
\subsection*{Scattering form and Stokes polytopes for the $n=10$ case}
\begin{figure}[H] 
\centering
\includegraphics[width=12cm]{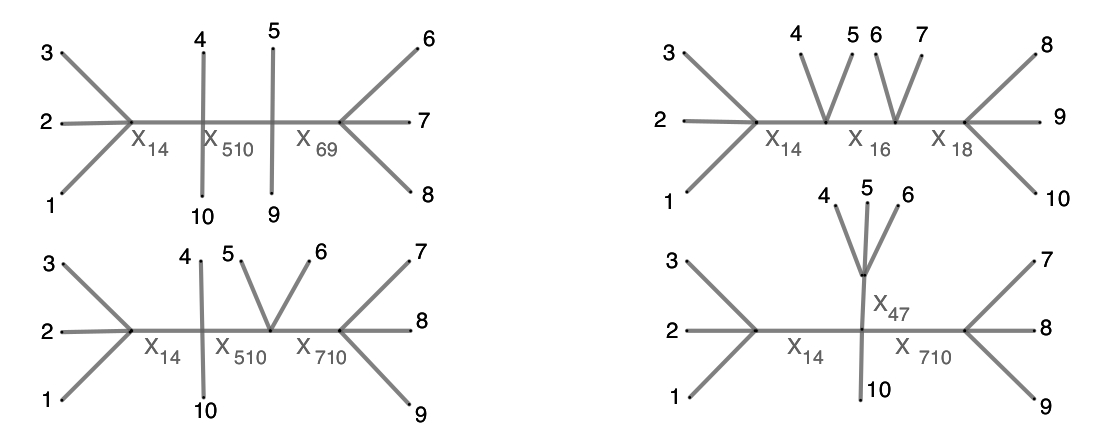}
\caption{The primitive quartic graphs (in clockwise order) with corresponding Stokes polytopes being Cube, Associahedron(2-4), Lucas and Mixed Classes(6 and 7)}
\label{fig: quarticgraphs}
\end{figure}

We would like to provide the details of how to obtain the Scattering amplitude $ \mathcal{M}_{10} $ by summing over  the kinematic Stokes polytopes here. There are a total of $F_4 = 55$ quadrangulations the sum over all of them can equivalently be replaced with a sum over just the $7$ primitive Stokes polytopes corresponding to the quartic graphs shown below (\ref{fig: quarticgraphs}) with appropriate coefficients. The reference quadrangulations for these primitves are $Q_{1}= (14,510,69), \; Q_{2}= (14,16,18), \; Q_{3}= (14,16,69), \; Q_{4}= (14,49,69),\; Q_{5}=(14,47,710),\; Q_{6}=(14,510,710), \; Q_{7}= (14,16,710)$ 
\vspace{0.5 cm}
We first provide the details of these Stokes polytopes and demonstrate how to get the planar scattering form, which when pulled back gives the scattering amplitude.\\
We always impose the associahedron condtions: 

\bea \label{associa}
s_{ij}=-c_{ij} \; \text{for} \; 1 \le i < j\le n-1 ,\;\; |i-j| \ge 2
\eea
and together with this we  need to impose $4$ additional conditions which carve out the Stokes polytope inside the associahedron. As explained in section \ref{locat=n6section} we consider the reference quadrangulation $Q$ corresponding to each Stokes polytope and find any set of 4 other diagonals $T$ that complete the triangulation of $Q$. There are 16 possible choices for such a set which correspond to choosing either of the two diagonals of each quadrilateral inside the reference quadrangulation independently. We choose any one of these sets. We then set the $X_{ij}$'s corresponding to this set to positive constants $d_{ij}$'s, since these $X_{ij}$'s can never correspond to propagators of any quartic graph.  This particular choice of additional contraints provides a particular embedding of the Stokes polytope into the associahedron.
We illustrate this for all the four cases below.
\begin{enumerate}
\item {\bf Cube type :} The corresponding Polytope is a cube with 8 vertices as shown in the figure \ref{fig: CUBE}.The set of $Q_{1}$ compatible quadrangulations are given by:\\
\begin{align}
S_{1} = \{ (14,510,69,+), \;(310,510,69,-),\; (14,49,69,-),\; (14,510,58,-),\nonumber \\ (14,49,58,+),(310,510,58,+),(310,49,69,+),\; (310,49,58,-) \} \nonumber
\end{align}
\begin{figure}[H]
\centering
\includegraphics[width=10cm]{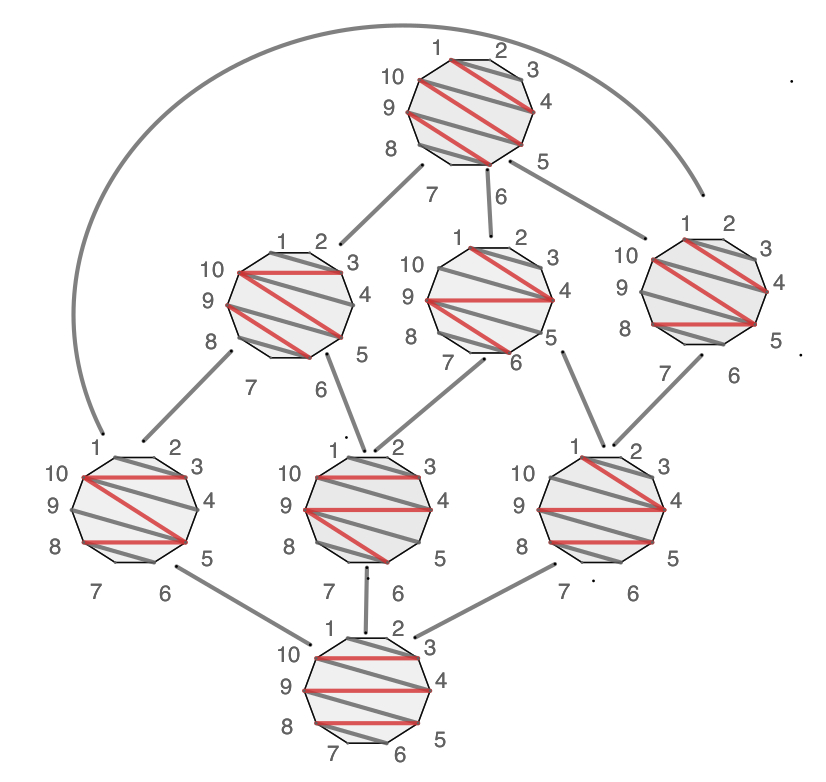}
\caption{\label{fig: CUBE}The Polytope is a cube as can been seen above each quadrangulation is a vertex and the lines joining them represent edges, each closed loop represents a face. The set of common diagonals which complete the triangulation are shown in grey.}
\end{figure}
\vspace{0.5 cm}
\noindent One set of diagonals which triangulate $Q_1$ are $T_1 = \{13,410,59,68\} $ which we set to positive constants to get an embedding\\
\bea \label{embedcu}
X_{13}=d_{13}, \;\; X_{410}=d_{410}, \;\; X_{59} =d_{59}, \;\; X_{68} =d_{68}
\eea
The planar scattering form for this case is given by:
\bea
\Omega^{Q_1}_{10}= (d\ln X_{14} \wedge d \ln X_{510}\wedge d \ln X_{69} -d\ln X_{310} \wedge d \ln X_{510}\wedge d \ln X_{69} \nonumber \\ -d\ln X_{14} \wedge d \ln X_{49}\wedge d \ln X_{69}- d\ln X_{310} \wedge d \ln X_{510}\wedge d \ln X_{58} + d\ln X_{14} \wedge d \ln X_{49}\wedge d \ln X_{58} \nonumber \\ + d\ln X_{310} \wedge d \ln X_{510}\wedge d \ln X_{58} + d\ln X_{310} \wedge d \ln X_{49}\wedge d \ln X_{69} -d\ln X_{310} \wedge d \ln X_{49}\wedge d \ln X_{58} ) \nonumber
\eea
When pulled back onto the space of constraints $\left( \ref{associa},\; \ref{embedcu} \right)$ gives the canonical form for the cube :\\
\bea
\begin{split}
 \omega^{Q_1}_{10} =  \Bigg( \Bigg. \frac{1}{X_{14} X_{510}X_{69}} + \frac{1}{X_{310} X_{510}X_{69}}+ \frac{1}{X_{14} X_{49 }X_{69}}+ \frac{1}{X_{14} X_{510}X_{58}} +\frac{1}{X_{14} X_{49}X_{58}} \nonumber \\ + \frac{1}{X_{310} X_{510}X_{58}} +\frac{1}{X_{310} X_{49} X_{69}} +\frac{1}{X_{310} X_{49}X_{58}} \Bigg . \Bigg) dX_{14} \wedge dX_{510}\wedge dX_{69}
 \end{split}
 \eea
\item {\bf Snake type :} The corresponding polytope is an associahedron $\mathcal{A}_{6}$ with 14 vertices (see figure \ref{fig: Snake}). As Explained above there are three quadrangulations that correspond to this case namely  $Q_{2}= (14,16,18), \; Q_{3}= (14,16,69), \; Q_{4}= (14,49,69)$. We show how to get the planar scattering form and canonical form for $Q_2$ below :

\begin{figure}[H] 
	\centering
	\includegraphics[width=11cm]{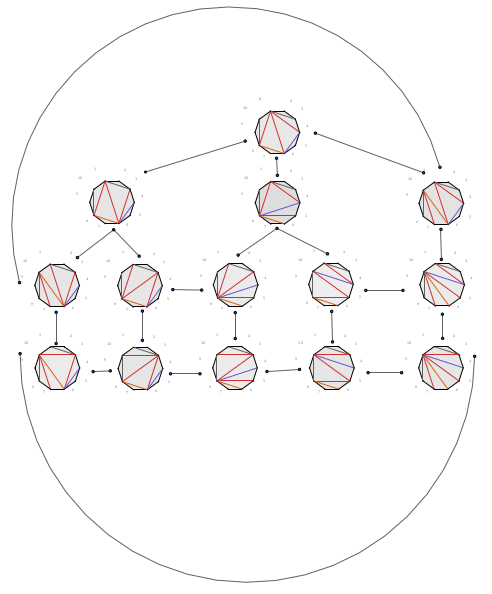}
	\caption{\label{fig: Snake}In the Snake case the corresponding Stokes polytope is an associahedron $\mathcal{A}_{6}$.}
\end{figure}
\vspace{0.5 cm}

The set of $Q_2$ compatible quadrangulations are given by:
\hspace{-1 cm}
\bea
S_2 = \{ (14,16,18,+),(36,16,18,-),(14,58,18,-),(14,16,710,-),(36,16,710,+),(36,38,18,+) \nonumber \\ ,(14,58,510,+),(38,58,18,+),(14,510,710,+),(36,310,710,-),(36,38,310,-),(310,58,510,-)\nonumber\\ 
,(38,58,310,-),(310,510,710,-) \}\nonumber
\eea
One set of diagonals which triangulates the reference quadrangulation $Q_{2}$ is $ T_2 = \{13,46, 68,810\}$ which we set to positive constants to get an embedding: 
 \bea \label{embedsnake}
 X_{13} =d_{13} \;, \; X_{46}=d_{46} \;,\; X_{68}=d_{68} \; , \; X_{810} =d_{810} 
\eea
The planar scattering form for this case is given by, 
\bea
\Omega^{Q_2}_{10}=  d\ln X_{14} \wedge d \ln X_{16}\wedge d \ln X_{18} -d\ln X_{36} \wedge d \ln X_{16}\wedge d \ln X_{18} \nonumber \\ -d\ln X_{14} \wedge d \ln X_{58}\wedge d \ln X_{18}- d\ln X_{14} \wedge d \ln X_{16}\wedge d \ln X_{710} + d\ln X_{36} \wedge d \ln X_{16}\wedge d \ln X_{710} \nonumber \\ + d\ln X_{36} \wedge d \ln X_{38}\wedge d \ln X_{18} + d\ln X_{14} \wedge d \ln X_{58}\wedge d \ln X_{510} + d\ln X_{38} \wedge d \ln X_{58}\wedge d \ln X_{18} \nonumber \\ + d\ln X_{14} \wedge d \ln X_{510}\wedge d \ln X_{710} - d\ln X_{36} \wedge d \ln X_{310}\wedge d \ln X_{710} -d\ln X_{36} \wedge d \ln X_{38}\wedge d \ln X_{310} \nonumber \\ - d\ln X_{310} \wedge d \ln X_{58}\wedge d \ln X_{510} - d\ln X_{38} \wedge d \ln X_{58}\wedge d \ln X_{310} - d\ln X_{310} \wedge d \ln X_{510}\wedge d \ln X_{710} \nonumber
\eea
When pulled back onto the space of constraints eqn. \eqref{associa} and eqn. \eqref{embedsnake} we get the canonical form:
\bea
\begin{split}
 \omega^{Q_2}_{10} = \Bigg( \Bigg. \frac{1}{X_{14} X_{16}X_{18}} + \frac{1}{X_{36} X_{16}X_{18}}+ \frac{1}{X_{14} X_{58}X_{18}}+ \frac{1}{X_{14} X_{16} X_{710}} + \frac{1}{X_{36} X_{16}X_{710}} \nonumber \\ + \frac{1}{X_{36}X_{38}X_{18}} +\frac{1}{X_{14} X_{58} X_{510}} + \frac{1}{X_{38} X_{58}X_{18}} + \frac{1}{X_{14} X_{510}X_{710}} + \frac{1}{X_{36} X_{310}X_{710}} \nonumber \\ + \frac{1}{X_{36} X_{38}X_{310}}+ \frac{1}{X_{310} X_{58} X_{510}} +\frac{1}{X_{38} X_{58}X_{310}}+ \frac{1}{X_{310} X_{510}X_{710}} \Bigg . \Bigg) dX_{14} \wedge dX_{16}\wedge dX_{18}
 \end{split}
 \eea
 Similarly,
 \bea
\begin{split}
 \omega^{Q_3}_{10} = \Bigg( \Bigg. \frac{1}{X_{14} X_{49}X_{69}} + \frac{1}{X_{310} X_{49}X_{69}}+ \frac{1}{X_{14} X_{16}X_{69}}+ \frac{1}{X_{14} X_{49} X_{58}} + \frac{1}{X_{36} X_{310}X_{69}} \nonumber \\ + \frac{1}{X_{310}X_{49}X_{58}} +\frac{1}{X_{16} X_{36} X_{69}} + \frac{1}{X_{14} X_{16}X_{18}} + \frac{1}{X_{14} X_{18}X_{58}} + \frac{1}{X_{36} X_{38}X_{310}} \nonumber \\ + \frac{1}{X_{38} X_{310}X_{58}}+ \frac{1}{X_{16} X_{18} X_{36}} +\frac{1}{X_{18} X_{38}X_{58}}+ \frac{1}{X_{18} X_{36}X_{38}} \Bigg . \Bigg) dX_{14} \wedge dX_{16}\wedge dX_{18}
 \end{split}
 \eea
 \bea
\begin{split}
 \omega^{Q_4}_{10} = \Bigg( \Bigg. \frac{1}{X_{14} X_{16}X_{69}} + \frac{1}{X_{16} X_{36}X_{69}}+ \frac{1}{X_{14} X_{510}X_{69}}+ \frac{1}{X_{14} X_{16} X_{18}} + \frac{1}{X_{16} X_{18}X_{36}} \nonumber \\ + \frac{1}{X_{36}X_{310}X_{69}} +\frac{1}{X_{310} X_{510} X_{69}} + \frac{1}{X_{14} X_{58}X_{510}} + \frac{1}{X_{14} X_{18}X_{58}} + \frac{1}{X_{18} X_{36}X_{38}} \nonumber \\ + \frac{1}{X_{36} X_{38}X_{310}}+ \frac{1}{X_{310} X_{58} X_{510}} +\frac{1}{X_{18} X_{38}X_{58}}+ \frac{1}{X_{38} X_{310}X_{58}} \Bigg . \Bigg) dX_{14} \wedge dX_{16}\wedge dX_{18}
 \end{split}
 \eea
\item {\bf Lucas type :} 
In this the corresponding Stokes polytope has Lucas number $L_{3}=12$ vertices (see figure \ref{fig: LUCAS}).
\begin{figure}[H]
\centering
\includegraphics[width=11cm]{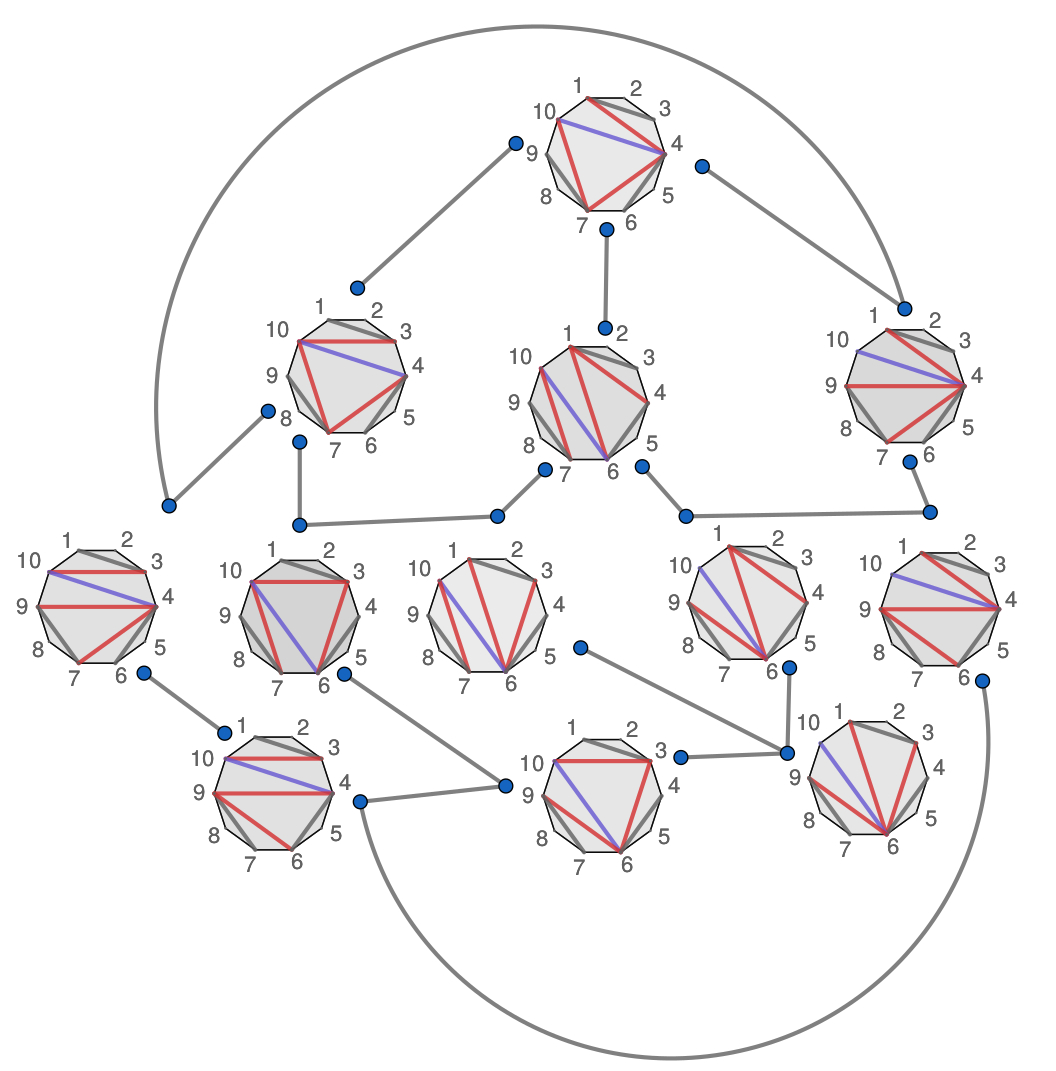}
\caption{\label{fig: LUCAS}In the Lucas case the corresponding polytope has 12 vertices, 18 edges and 8 faces.}
\end{figure}
The set of $Q_5$ compatible quadrangulations are given by:
\hspace{-1 cm}
\bea
S_5 = \{ (14,47,710,+),(310,47,710,-),(14,16,710,-),(14,47,49,-),(310,49,47,+),(36,310,710,+) \nonumber \\ ,(36,16,710,+),(14,16,69,+),(14,49,69,+),(310,49,69,-),(310,36,69,-),(36,16,69,-) \}\nonumber
\eea
One set of diagonals which triangulates the reference quadrangulation $Q_{5}$ is $T_3 = \{13,46, 79,410\}$ which we set to positive constants to get an embedding: 
 \bea \label{embedlucas}
 X_{13} =d_{13} \;, \; X_{46}=d_{46} \;,\; X_{79}=d_{79} \; , \; X_{410} =d_{410} 
\eea
The planar scattering form for this case is given by,

\bea
\Omega^{Q_5}_{10}= d\ln X_{14} \wedge d \ln X_{47}\wedge d \ln X_{710} -d\ln X_{310} \wedge d \ln X_{47}\wedge d \ln X_{710} \nonumber \\ -d\ln X_{14} \wedge d \ln X_{16}\wedge d \ln X_{710}- d\ln X_{14} \wedge d \ln X_{47}\wedge d \ln X_{49} + d\ln X_{310} \wedge d \ln X_{49}\wedge d \ln X_{47} \nonumber \\ + d\ln X_{36} \wedge d \ln X_{310}\wedge d \ln X_{710} + d\ln X_{36} \wedge d \ln X_{16}\wedge d \ln X_{710} + d\ln X_{14} \wedge d \ln X_{16}\wedge d \ln X_{69} \nonumber \\ + d\ln X_{14} \wedge d \ln X_{49}\wedge d \ln X_{69} - d\ln X_{310} \wedge d \ln X_{49}\wedge d \ln X_{69} -d\ln X_{310} \wedge d \ln X_{36}\wedge d \ln X_{69} \nonumber \\ - d\ln X_{36} \wedge d \ln X_{16}\wedge d \ln X_{69} \nonumber
\eea
When pulled back onto the space of constraints eqn. \eqref{associa} and eqn. \eqref{embedlucas} we get the canonical form:
\bea
\begin{split}
 \omega^{Q_5}_{10} = \Bigg( \Bigg. \frac{1}{X_{14} X_{47}X_{710}} + \frac{1}{X_{310} X_{47}X_{710}}+ \frac{1}{X_{14} X_{16}X_{710}}+ \frac{1}{X_{14} X_{47} X_{49}} + \frac{1}{X_{310} X_{49}X_{47}} \nonumber \\ + \frac{1}{X_{36}X_{310}X_{710}} +\frac{1}{X_{36} X_{16} X_{710}} + \frac{1}{X_{14} X_{16}X_{69}} + \frac{1}{X_{14} X_{49}X_{69}} + \frac{1}{X_{310} X_{49}X_{69}} \nonumber \\ + \frac{1}{X_{310}X_{36}X_{69}}+ \frac{1}{X_{36} X_{16} X_{69}} \Bigg . \Bigg) dX_{14} \wedge dX_{47}\wedge dX_{710}
 \end{split}
 \eea
\item {\bf Mixed type :} In this case the stokes polytope is just product of lower dimensional stokes polytopes $S_{1} \times S_{2_2}$ hence has 10 vertices (see figure \ref{fig: MIXED}). As explained above there are two quadrangulations that correspond to this case namely  $Q_{6}=(14,510,710), \; Q_{7}= (14,16,710)$. We show how to get the planar scattering form and canonical form for $Q_6$ below:

The set of $Q_6$ compatible quadrangulations are given by:
\hspace{-1 cm}
\bea
S_6 = \{ (14,510,710,+),(310,510,710,-),(14,47,710,-),(14,510,69,-),(310,47,710,+), \nonumber \\(310,510,69,+),(14,47,49,+),(14,49,69,+),(310,49,69,-),(310,47,49,-) \}\nonumber
\eea
One set of diagonals which triangulates the reference quadrangulation $Q_{6}$ is $T_6 = \{13,410, 79,57 \}$ which we set to positive constants to get an embedding: 
 \bea \label{embedmixed}
 X_{13} =d_{13} \;, \; X_{410}=d_{410} \;,\; X_{79}=d_{79} \; , \; X_{57} =d_{57} 
\eea

\begin{figure}[H]
	\centering
	\includegraphics[width=10cm]{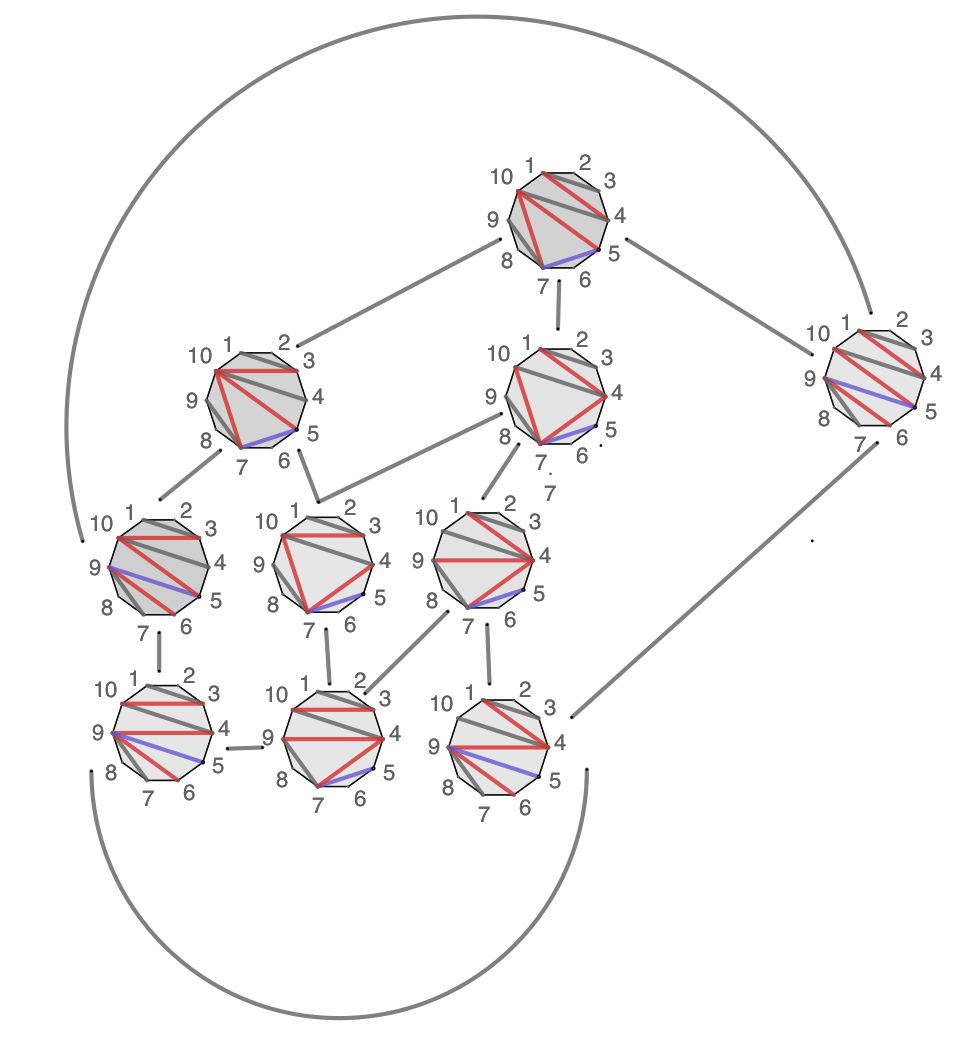}
	\caption{\label{fig: MIXED}In the mixed case the corresponding polytope has 10 vertices, 15 edges and 7 faces.}
\end{figure}  

The planar scattering form for this case is,
\bea
\Omega^{Q_6}_{10}= ( d\ln X_{14} \wedge d \ln X_{510}\wedge d \ln X_{710} -d\ln X_{310} \wedge d \ln X_{510}\wedge d \ln X_{710} \nonumber \\ -d\ln X_{14} \wedge d \ln X_{47}\wedge d \ln X_{710}- d\ln X_{14} \wedge d \ln X_{510}\wedge d \ln X_{69} + d\ln X_{310} \wedge d \ln X_{47}\wedge d \ln X_{710} \nonumber \\ + d\ln X_{310} \wedge d \ln X_{510}\wedge d \ln X_{69} + d\ln X_{14} \wedge d \ln X_{47}\wedge d \ln X_{49} + d\ln X_{14} \wedge d \ln X_{49}\wedge d \ln X_{69} \nonumber \\ - d\ln X_{310} \wedge d \ln X_{49}\wedge d \ln X_{69} - d\ln X_{310} \wedge d \ln X_{47}\wedge d \ln X_{49}) \nonumber
\eea
When pulled back onto the space of constraints (\ref{associa},\ref{embedmixed}) we get the canonical form:
\bea
\begin{split}
 \omega^{Q_6}_{10} = \Bigg( \Bigg. \frac{1}{X_{14} X_{510}X_{710}} + \frac{1}{X_{310} X_{510} X_{710}}+ \frac{1}{X_{14} X_{47}X_{710}}+ \frac{1}{X_{14} X_{510} X_{69}} + \frac{1}{X_{310} X_{47}X_{710}} \nonumber \\ + \frac{1}{X_{310}X_{510}X_{69}} +\frac{1}{X_{14} X_{47} X_{49}} + \frac{1}{X_{14} X_{49}X_{69}} + \frac{1}{X_{310} X_{49}X_{69}} + \frac{1}{X_{310} X_{47}X_{49}}  \Bigg . \Bigg) \nonumber \\ dX_{14} \wedge dX_{510}\wedge dX_{710}
 \end{split}
 \eea
 Similarly,
 \bea
\begin{split}
 \omega^{Q_7}_{10} = \Bigg( \Bigg. \frac{1}{X_{14} X_{16}X_{710}} + \frac{1}{X_{16} X_{36} X_{710}}+ \frac{1}{X_{14} X_{510}X_{710}}+ \frac{1}{X_{14} X_{16} X_{69}} + \frac{1}{X_{36} X_{310}X_{710}} \nonumber \\ + \frac{1}{X_{16}X_{36}X_{69}} +\frac{1}{X_{310} X_{510} X_{710}} + \frac{1}{X_{14} X_{510}X_{69}} + \frac{1}{X_{36} X_{310}X_{69}} + \frac{1}{X_{310} X_{510}X_{69}}  \Bigg . \Bigg) \nonumber \\ dX_{14} \wedge dX_{16}\wedge dX_{710}
 \end{split}
 \eea
 
\end{enumerate}
Upon substituting the corresponding $m_{10}$ in eqn.(8), it can be checked that for $\alpha_{Q_1}= \frac{5}{24}$, $\alpha_{Q_2}= \alpha_{Q_3}=\alpha_{Q_4}=\frac{1}{24}$, $\alpha_{Q_5}= \frac{2}{24}$ and $\alpha_{Q_6}= \alpha_{Q_7}=\frac{3}{24}$ the sum over all the residues give $ \mathcal {M}_{10}$.

\bibliographystyle{bibstyle}
\bibliography{Stokes}
\end{document}